# Virtual Machine Warmup Blows Hot and Cold*

EDD BARRETT, King's College London, UK
CARL FRIEDRICH BOLZ-TEREICK, King's College London, UK
REBECCA KILLICK, Lancaster University, UK
SARAH MOUNT, King's College London, UK
LAURENCE TRATT, King's College London, UK

Virtual Machines (VMs) with Just-In-Time (JIT) compilers are traditionally thought to execute programs in two phases: the initial warmup phase determines which parts of a program would most benefit from dynamic compilation, before JIT compiling those parts into machine code; subsequently the program is said to be at a steady state of peak performance. Measurement methodologies almost always discard data collected during the warmup phase such that reported measurements focus entirely on peak performance. We introduce a fully automated statistical approach, based on changepoint analysis, which allows us to determine if a program has reached a steady state and, if so, whether that represents peak performance or not. Using this, we show that even when run in the most controlled of circumstances, small, deterministic, widely studied microbenchmarks often fail to reach a steady state of peak performance on a variety of common VMs. Repeating our experiment on 3 different machines, we found that at most 43.5% of ⟨VM, benchmark⟩ pairs consistently reach a steady state of peak performance.

## 1 INTRODUCTION

Many modern languages are implemented as Virtual Machines (VMs) which use a Just-In-Time (JIT) compiler to translate 'hot' parts of a program into efficient machine code at run-time. Since it takes time to identify and JIT compile the 'hot' parts of a program, VMs using a JIT compiler are said to be subject to a *warmup* phase. The traditional view of JIT compiled VMs is that program execution is slow during the warmup phase, and fast afterwards, when a *steady state* of *peak performance* is said to have been reached. This traditional view underlies nearly all JIT compiler benchmarking methodologies: after running benchmarks many times within a single VM process,

*Updates to this paper will be found at https://arxiv.org/abs/1602.00602

Authors' URLs: E. Barrett http://eddbarrett.co.uk/, C. F. Bolz-Tereick http://cfbolz.de/, R. Killick http://www.lancs.ac.uk/~killick/, S. Mount http://snim2.org/, L. Tratt http://tratt.net/laurie/.
Authors' addresses: Edd Barrett, King's College London, UK; Carl Friedrich Bolz-Tereick, King's College London, UK; Rebecca Killick, Lancaster University, UK; Sarah Mount, King's College London, UK; Laurence Tratt, King's College London, UK.

the data collected before warmup is discarded, with the reported measurements focussing only on peak performance.

The fundamental aim of this paper is to test, in a highly idealised setting, the following hypothesis:

**H1** Small, deterministic programs reach a steady state of peak performance.

To test this hypothesis, we developed a new approach for automatically analysing large quantities of benchmarking data. We deliberately chose widely studied microbenchmarks, and ran them in a heavily controlled environment, designed to minimise measurement noise. By doing so, we maximised the chances that our experiment would validate Hypothesis H1. Instead we encountered many benchmarks which slowdown, never hit a steady state, or have inconsistent performance from one run to the next. Only 30.0–43.5% (depending on the machine and OS combination) of ⟨VM, benchmark⟩ pairs consistently reach a steady state of peak performance. Of the seven VMs we studied, none consistently reached a steady state of peak performance. These results are much worse than reported in previous works.

Our results suggest that much real-world VM benchmarking, which nearly all relies on assuming that benchmarks do reach a steady state of peak performance, is likely to be partly or wholly misleading. Since microbenchmarks similar to those in this paper are often used in isolation to gauge the efficacy of VM optimisations, it is also likely that ineffective, or deleterious, optimisations may have been incorrectly judged as improving performance and included in VMs.

### 1.1 Overview of the Methodology

We present a carefully designed experiment where benchmarks are run for 2000 *in-process iterations* and repeated using 30 fresh *process executions* (i.e. each process execution runs multiple in-process iterations). We then automatically analyse and classify the resulting data.

In order to reduce the influence of external factors, we wrote a new benchmarking tool, Krun, to control as many confounding variables as practical. For example, Krun reboots machines before each process execution, turns off or stops network cards and userland daemons, and ensures that the machine's temperature is consistent before each process execution is run. We ran our experiment on different machines and operating systems to understand the effect of both on benchmarks.

With the time series data produced by Krun, we then encounter the issue of producing useful and reliable summary statistics. Traditional VM benchmarking uses simple heuristics to guess when warmup has completed, e.g. Georges et al. [2007]. Kalibera and Jones [2013] convincingly show the limitations of such approaches, presenting instead a manual approach to determining if and when a steady state has been reached. While this is a significant improvement on previous methods, it is time-consuming, prone to human inconsistency, and gives no indication as to whether the steady state represents peak performance or not.

We use changepoint analysis [Eckley et al. 2011] to automatically analyse VM benchmarking data. Changepoint analysis detects shifts in the nature of time series data (e.g. when a benchmark switches from 'slow' to 'fast' modes after JIT compilation). We then use this information to automatically classify each process execution as either: no steady state; flat (no detectable change in performance over the benchmark); warmup; or slowdown (a decrease in performance over time). Figure 1 shows an example of how changepoint analysis can automatically classify both 'good' and 'bad' warmup styles. Although simple, these classifications allow us to tease apart unexpected performance patterns that previous methodologies cannot.

Finally, in an attempt to see if there are easy explanations for the odd effects we see, we take two VMs and separate out the time taken to perform Garbage Collection (GC) and JIT compilation. These two factors explain only some of the effects we see.

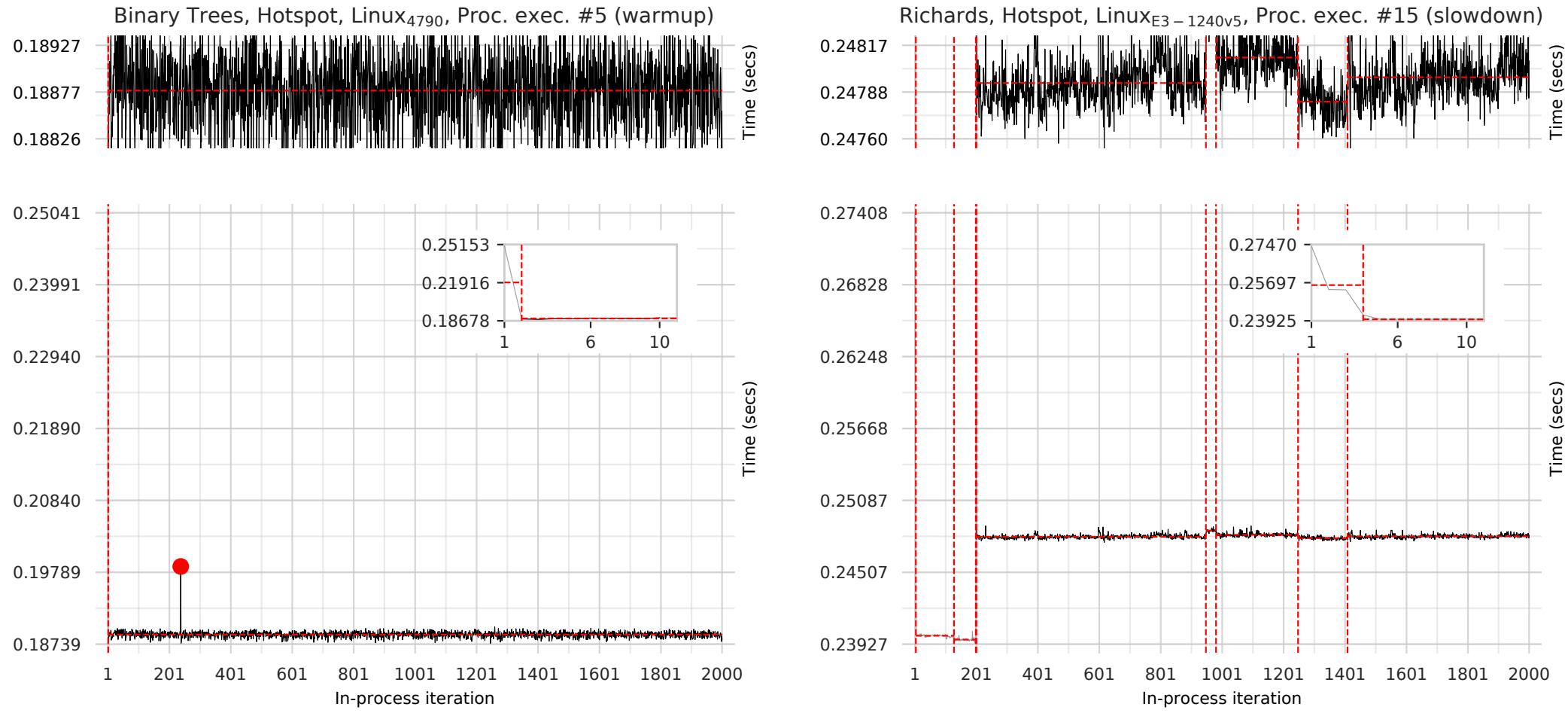

Fig. 1. Two example run-sequence plots from our results, augmented with the results of changepoint analysis. The LHS plot shows an example of 'traditional' warmup (in this case *binary trees* on HotSpot); the RHS plot shows an example of slowdown (*Richards* on HotSpot). The plot titles show: the benchmark (e.g. *binary trees*), the VM (e.g. HotSpot), the machine (e.g. $\text{Linux}_{4790}$), the process execution number (e.g. 5), and the classification (e.g. warmup). The 'main' plot at the bottom always shows the in-process iteration number (indexed from 1) and the corresponding wall-clock times for all 2000 in-process iterations. Because this often obscures the finer details, we also show two additional visualisations of the data: the top plot shares its $x$ axis with the bottom plot, but zooms the $y$ axis in to where the bulk of the in-process timing data is located; the smaller inset plot allows us to examine a handful of in-process iteration timings (typically, though not always, the first $n$, allowing us to see the 'warmup curve', if it exists). Changepoints are indicated by dashed vertical red lines and denote a 'shift' in performance behaviour; changepoint segments are indicated by dashed horizontal red lines between changepoints, whose height shows the mean iteration time within the segment; and red circles indicate 'outliers', which the changepoint analysis discount. We use the changepoint segments to determine if and when a steady state has been reached and, if so, to classify the plots' warmup style (in this example, warmup and slowdown respectively). As an additional visual aid, changepoint segments considered equivalent in timing to the steady state segment are plotted in black; all other segments are plotted in grey.

### 1.2 Summary

In summary, this paper's contributions are as follows:

(1) We show that, even in a highly idealised setting, widely studied benchmarks often fail to reach a steady state of peak performance.
(2) We present the first automated approach to analysing time series data from VM benchmarking, which can automatically determine if and when a steady state has been reached.
(3) We present the first classification of different styles of steady state behaviour (warmup, slowdown, flat).
(4) We show that, while some instances of 'odd' behaviour are explained by the 'obvious' factors of GC and JIT compilation, many are not.

The statistical approach we have developed is open-source, general, and can be used to analyse any VM benchmarking data. In the Appendix, we show it applied to the DaCapo and Octane benchmark suites, both run in a conventional manner (Appendix A). We also present a curated series of plots of interesting data (Appendix C). A separate document contains the complete plots of all data from all machines.

Our repeatable experiment, as well as the specific data that forms the basis of this paper's results, can be downloaded from:

https://archive.org/download/softdev_warmup_experiment_artefacts/v0.8/

## 2 BACKGROUND

When a program begins running on a JIT compiling VM, it is typically (slowly) interpreted; once 'hot' (i.e. frequently executed) loops or methods are identified, they are dynamically compiled into machine code; and subsequent executions of those loops or methods use (fast) machine code rather than the (slow) interpreter. Once machine code generation has completed, the VM is said to have finished warming up, and the program is said to be executing at a steady state of peak performance.[1] While the length of the warmup period is dependent on the program and JIT compiler, all JIT compiling VMs are based on this performance model [Kalibera and Jones 2013].

Benchmarking of JIT compiled VMs typically focusses on reporting steady state numbers based on two assumptions: first, that warmup is both fast and inconsequential to users; second, that the steady state is also peak performance. The methodologies used are typically straightforward: benchmarks are run for a number of in-process iterations within a single VM process execution. The first $n$ in-process iterations are then discarded, on the basis that warmup *should* have completed in that period. Often $n$ is a fixed number (typically around 5), with no guarantee that warmup has completed by that point. More advanced methods such as that of Georges et al. [2007] try to find a steady state using simple statistical tests.

The most sophisticated VM benchmarking analysis yet developed is found in [Kalibera and Jones 2012, 2013]. After showing that simple heuristics such as Georges et al.'s often fail to accurately pinpoint when warmup has completed, Kalibera and Jones then present an alternative manual process. In essence, after a specific VM / benchmark combination has been run for a small number of process executions, a human must determine if a steady state is reached; and, if it is, at which in-process iteration it is reached. When a steady state does exist, a larger number of process executions are then run; the previously determined cut-off point is applied to each process execution's in-process iterations; and detailed statistics are produced.

The Kalibera and Jones methodology is a significant advance on previous work, and is an important inspiration for ours. However, the reliance on manually identifying when warmup is complete has ramifications. Most obviously, humans are prone to error and disagreement when performing such identifications. More significantly, the time required to manually examine the time series data means that it is only practical to apply it to a small number of initial process executions: the cut-off point that is determined from those is then applied to all future process executions, even though there is no guarantee that they all reach a steady state by that point. With the exception of in-process iterations which show dependence in their data (e.g. a repeated 'cyclic' pattern), which are dealt with specially, Kalibera and Jones do not otherwise classify steady state performance relative to what came before it, making it hard to understand if the steady state represents peak performance or not. These three points mean that, despite Kalibera and Jones's advances, "determining when a system has warmed up, or even providing a rigorous definition of the term, is an open research problem" [Seaton 2015].

[1]The traditional view applies equally to VMs that perform immediate compilation instead of using an interpreter, and to those VMs which have more than one layer of JIT compilation (later JIT compilation is used for 'very hot' portions of a program, trading slower compilation time for better machine code generation).

## 3 METHODOLOGY

To test Hypothesis H1, we designed an experiment which uses a suite of microbenchmarks: each is run with 2000 in-process iterations and repeated using 30 process executions. We have carefully designed our experiment to be repeatable and to control as many potentially confounding variables as is practical. In this section we detail: the benchmarks we used and the modifications we applied; the VMs we benchmarked; the machines we used for benchmarking; and the Krun system we developed to run benchmarks.

First time readers of this paper may find it easiest to jump straight to Section 4, coming back to the complete (lengthy!) methodology on a second read.

### 3.1 The Microbenchmarks

The microbenchmarks we use are as follows: *binary trees*, *spectralnorm*, *n-body*, *fasta*, and *fannkuch redux* from the Computer Language Benchmarks Game (CLBG) [Bagley et al. 2004]; and *Richards*. Readers can be forgiven for initial scepticism about this set of microbenchmarks. They are small and widely used by VM authors as optimisation targets. In general they are more effectively optimised by VMs than average programs; when used as a proxy for other types of programs (e.g. large programs), they tend to overstate the effectiveness of VM optimisations (see e.g. [Ratanaworabhan et al. 2009]). In our context, this weakness is in fact a strength: small, deterministic, and widely examined programs are our most reliable means of testing Hypothesis H1. Put another way, if we were to run arbitrary programs and find unusual warmup behaviour, a VM author might reasonably counter that "you have found the one program that exhibits unusual warmup behaviour".

For each benchmark, we provide versions in C, Java, JavaScript, Python, Lua, PHP, and Ruby. Since most of these benchmarks have multiple implementations in any given language, we picked the versions used in [Bolz and Tratt 2015], which represented the fastest performers at the point of that publication. We lightly modified the benchmarks to integrate with our benchmark runner (see Section 3.5) but did not e.g. force garbage collection.

*3.1.1 Ensuring Determinism.* User programs that are deliberately non-deterministic are unlikely to warm-up in the traditional fashion. We therefore wish to guarantee that our benchmarks are, to the extent controllable by the user, deterministic, taking the same path through the Control Flow Graph (CFG) on all process executions and in-process iterations. We make no attempt to control non-determinism within the VM, which is part of what we need to test for Hypothesis H1.

To check whether the benchmarks were deterministic at the user-level, we created versions with `print` statements at all possible points of CFG divergence (e.g. inside conditional branches). These versions are available in our experimental suite. We first ran the modified benchmarks with 2 process executions and 20 in-process iterations, and compared the outputs of the two processes. This was enough to show that the *fasta* benchmark was non-deterministic in all language variants, due to its random number generator not being reseeded. We fixed this by moving the random seed initialisation to the start of the in-process iteration main loop.

In order to understand the effects of compilation non-determinism, we then compiled the VMs and ran our modified benchmarks on two different machines. We then observed occasional non-determinism in Java benchmarks. This was due to the extra class we had added to each benchmark to interface with the benchmark runner: sometimes, the main benchmark class was lazily loaded after benchmark timing had started in a way that we could observe. We solved this by adding an empty static method to each benchmark, which our extra classes then call via a static initialiser, guaranteeing that the main benchmark class is eagerly loaded. Note that we do not attempt to eagerly load other classes: lazy loading is an inherent part of the JVM specification, and thus something that should be part of our measurements.

### 3.2 Measuring Computation Rather than File Performance

By their very nature, microbenchmarks tend to perform computations which can be easily optimised away. While this speaks well of optimising compilers, benchmarks whose computations are entirely removed are rarely useful [Seaton 2015]. To prevent optimisers removing such code, many benchmarks write intermediate and final results to `stdout`. However, this then means that one starts including the performance of file routines in libraries and the kernel in measurements, which can become a significant part of the eventual measure.

To avoid this, we modified the benchmarks to calculate a checksum during each in-process iteration. At the end of each in-process iteration the checksum is compared to an expected value; if the comparison fails then the incorrect checksum is written to `stdout`. This idiom lowers the chances that an optimiser can remove the main benchmark code, even though no output is produced. We also use this mechanism to give some assurance that the different language implementations of a benchmark are performing roughly the same work, as the expected checksum value is the same for all implementations of the benchmark.

### 3.3 VMs under Investigation

We ran the benchmarks on the following language implementations: GCC 4.9.3; Graal 0.18 (an alternative JIT compiler for HotSpot); HHVM 3.15.3 (a JIT compiling VM for PHP); JRuby+Truffle 9.1.2.0 (a JIT compiling VM for Ruby using Graal/Truffle); HotSpot 8u112b15 (the most widely used Java VM); LuaJIT 2.0.4 (a tracing JIT compiling VM for Lua); PyPy 5.6.0 (a meta-tracing JIT compiling VM for Python 2.7); and V8 5.4.500.43 (a JIT compiling VM for JavaScript). A repeatable build script downloads, patches, and builds fixed versions of each VM. All VMs were compiled with GCC/G++ 4.9.3 (and GCC/G++ bootstraps itself, so that the version we use compiled itself) to remove the possibility of variance through the use of different compilers.

We skipped: Graal, HHVM, and JRuby+Truffle on OpenBSD, as these VMs have not yet been ported to this platform; *fasta* on JRuby+Truffle as it crashes; and *Richards* on HHVM since it takes as long as every other benchmark on every other VM put together.

### 3.4 Benchmarking Hardware

With regards to hardware and operating systems, we made the following hypothesis:

**H2** Moderately different hardware and operating systems have little effect on warmup.

We deliberately use the word ‘moderately’, since significant changes of hardware (e.g. x86 vs. ARM) or operating system (e.g. Linux vs. Windows) imply that significantly different parts of the VMs will be used (see Section 7).

In order to test Hypothesis H2, we used three benchmarking machines: $\mathit{Linux}_{\mathrm{1240v5}}$, a Xeon E3-1240 v5 3.5GHz, 24GB of RAM, running Debian 8; $\mathit{Linux}_{4790}$, a quad-core i7-4790 3.6GHz, 32GB of RAM, running Debian 8; and $\mathit{OpenBSD}_{4790}$, a quad-core i7-4790 3.6GHz, 32GB of RAM, running OpenBSD 6.0. $\mathrm{Linux}_{\mathrm{1240v5}}$ and $\mathrm{Linux}_{4790}$ have the same OS (with the same packages and updates etc.) but different hardware; $\mathrm{Linux}_{4790}$ and $\mathrm{OpenBSD}_{4790}$ have the same hardware (to the extent we can determine) but different operating systems.

We disabled turbo boost and hyper-threading in the BIOS. Turbo boost allows CPUs to temporarily run in an higher-performance mode; if the CPU deems it ineffective, or if its safe limits (e.g. temperature) are exceeded, turbo boost is reduced [Charles et al. 2009]. Turbo boost can thus substantially change one’s perception of performance. Hyper-threading gives the illusion that a single physical core is two logical cores. Programs and threads that would otherwise run on separate physical cores are thus inter-leaved on a single core, leading to less predictable performance.

### 3.5 Krun

Many confounding variables manifest shortly before, or during the running of, benchmarks [Kalibera et al. 2005]. In order to control as many of these as possible, we wrote Krun[2]. Krun is a 'supervisor' which, given a configuration file specifying VMs, benchmarks, etc. configures a Linux or OpenBSD system, runs benchmarks, and collects the results. Individual VMs and benchmarks are then wrapped, or altered, to report data back to Krun in an appropriate format.

In the remainder of this subsection, we describe Krun. Since most of Krun's controls work identically on Linux and OpenBSD, we start with those, before detailing the differences imposed by the two operating systems. We then describe how Krun collects data from benchmarks. Note that, although Krun has various 'developer' flags to aid development and debugging benchmarking suites, we describe only Krun's full 'production' mode.

*3.5.1 Platform Independent Controls.* A typical problem with benchmarking is that earlier process executions can affect later ones (e.g. a benchmark which forces memory to swap will make later benchmarks seem to run slower). Therefore, before each process execution (including before the first), Krun reboots the system, ensuring that each process execution runs with the machine in a state that is uninfluenced by previous process executions. After each reboot, Krun is executed by the init subsystem; Krun then pauses for 3 minutes to allow the system to fully initialise; calls `sync` (to flush any remaining files to disk) followed by a 30 second wait; before finally running the next process execution.

The obvious way for Krun to determine which benchmark to run next is to examine its results file. However, this is a large file which grows over time, and reading it in could affect benchmarks (e.g. due to significant memory fragmentation). When invoked for the first time, Krun creates a simple schedule file. After each reboot this is scanned line-by-line for the next benchmark to run; the benchmark is run; and the schedule updated, without changing its size. Once the process execution is complete, Krun can safely load the results file in and append the results data, knowing that the reboot that will occur shortly after will put the machine into a (largely) known state.

Modern systems have various temperature-based limiters built in: CPUs, for example, lower their frequency if they get too hot. After its initial invocation, Krun waits for 1 minute before collecting the values of all available temperature sensors. After each reboot's `sync`-wait, Krun waits for the machine to return to these base temperatures (±3°C) before starting the benchmark, fatally aborting if this temperature range is not met within 1 hour. In so doing, we aim to lessen the impact of ambient temperature changes.

Krun fixes the heap and stack `ulimit` for all VM processes (in our case, 2GiB heap and a 8MiB stack).[3] Benchmarks are run as the '`krun`' user, whose account and home directory are created afresh before each process execution to prevent cached files affecting benchmarking.

User-configurable commands can be run before and after each process execution. We disabled as many Unix daemons as possible (e.g. smtpd, crond) to lessen the effects of context switching. We also turned off network interfaces entirely, to prevent outside sources causing (potentially performance interfering) interrupts to be sent to the processor and kernel.

In order to identify problems with the machine itself, Krun monitors the system `dmesg` buffer for unexpected entries (known 'safe' entries are ignored), informing the user if any arise. We implemented this after noticing that one machine initially ear-marked for benchmarking occasionally overheated during benchmarking, with the only clue to this being a line in `dmesg`. We did not use this machine for our final benchmarking.

[2]The 'K' in Krun is a respectful tip of the hat to Kalibera and Jones.

[3]Note that Linux allows users to inspect these values, but also to allocate memory that exceeds them.

A process's environment size can cause measurement bias [Mytkowicz et al. 2009]. The diversity of VMs and platforms in our setup makes it impossible to set a unified environment size across all VMs and benchmarks. However, the `krun` user does not vary its environment (we recorded the environment seen by each process execution and verified their size), and we designed the experiment such that, for each machine, each ⟨VM, benchmark⟩ has a consistent environment size.

*3.5.2 Linux-specific Controls.* On Linux, Krun controls several additional factors, sometimes by checking that the user has correctly set controls which can only be set manually.

Krun uses `cpufreq-set` to set the CPU governor to `performance` mode (i.e. the highest non-overclocked frequency possible). To prevent the kernel overriding this setting, Krun verifies that the user has disabled Intel P-state support in the kernel by passing `intel_pstate=disable` as a kernel argument.

As standard, Linux interrupts ('ticks') each core `CONFIG_HZ` times per second (usually 250) to decide whether to perform a context switch. To avoid these repeated interruptions, Krun checks that it is running on a 'tickless' kernel [Linux 2013], which requires recompiling the kernel with the `CONFIG_NO_HZ_FULL_ALL` option set. Whilst the boot core still 'ticks', other cores only 'tick' if more than one runnable process is scheduled.

Similarly, Linux's `perf` profiler may interrupt cores up to 100,000 times a second. We became aware of `perf` when Krun's `dmesg` checks notified us that the kernel had decreased the sample-rate by 50% due to excessive overhead. While frequent sampling will clearly have some performance impact, adaptive sampling is even more troubling: a change of sample rate during a process execution could notably change performance. Although `perf` cannot be completely disabled, Krun sets it to sample at most once per second, minimising interruptions.

*3.5.3 OpenBSD-specific Controls.* Relative to Linux, OpenBSD exposes fewer controls to user. Nevertheless, there are two OpenBSD specific features in Krun. First, Krun sets CPU performance to maximum by invoking `apm -H` prior to running benchmarks (equivalent to Linux's `performance` mode). Second, Krun minimises the non-determinism in OpenBSD's malloc implementation, for example not requiring `realloc` to always reallocate memory to an entirely new location. The `malloc.conf` flags we use are `cfgrux`.

*3.5.4 The Iteration Runners.* To report timing data to Krun, we created an *iteration runner* for each language under investigation. These take the name of a specific benchmark and the desired number of in-process iterations, run the benchmark appropriately, and once it has completed, print the times to `stdout` for Krun to capture. For each in-process iteration we measure (on Linux and OpenBSD) the wall-clock time taken, and (Linux only) core cycle, APERF, and MPERF counters.

We use a monotonic wall-clock timer with sub-millisecond accuracy (`CLOCK_MONOTONIC_RAW` on Linux, and `CLOCK_MONOTONIC` on OpenBSD). Although wall-clock time is the only measure which really matters to users, it gives no insight into multi-threaded computations: on Linux we also record core-cycle counts using the `CPU_CLK_UNHALTED.CORE` counter to see what work each core is actually doing. In contrast, we use the ratio of APERF/MPERF deltas solely as a safety check that our wall-clock times are valid. The `IA32_APERF` counter increments at a rate proportional to the processor's current frequency; the `IA32_MPERF` counter increments at a fixed rate normalised to the CPU's base frequency. With an APERF/MPERF ratio of precisely 1, the processor is running at full speed; below 1 it is in a power-saving mode; and above 1, turbo boost is being used.

A deliberate design goal of the in-process iteration runners is to minimise timing noise and distortion from measurements. Since system calls can have a significant overhead (on Linux, calling functions such as `write` can evict as much as two thirds of an x86's L1 cache [Soares and Stumm 2010]), we avoid making any system calls other than those required to take measurements. We

```
void krun_measure(int mdata_idx) {
  struct krun_data *data = &(krun_mdata[mdata_idx]);
  if (mdata_idx == 0) { // start benchmark readings
    for (int core = 0; core < num_cores; core++) {
      data->aperf[core] = read_aperf(core);
      data->mperf[core] = read_mperf(core);
      data->core_cycles[core] = read_core_cycles(core);
    }
    data->wallclock = krun_clock_gettime_monotonic();
  } else {               // end benchmark readings
    data->wallclock = krun_clock_gettime_monotonic();
    for (int core = 0; core < num_cores; core++) {
      data->core_cycles[core] = read_core_cycles(core);
      data->aperf[core] = read_aperf(core);
      data->mperf[core] = read_mperf(core);
    }
  }
}
```

Listing 1. `krun_measure`: Measuring before (the `if` true branch) and after (the false branch) a benchmark. Since wall-clock time is the most important measure, it is innermost; since the APERF/MPERF counters are a sanity check, they are outermost. Note that the APERF/MPERF counters must be read in the same order before and after a benchmark.

```
wallclock_times = [0] * iters
for i in xrange(iters):
  # Start timed section
  krun_measure(0)
  # Call the benchmark
  bench_func(param)
  # End timed section
  krun_measure(1)
  wallclock_times[i] = \
    krun_get_wallclock(1) - \
    krun_get_wallclock(0)
js = { "wallclock_times": \
       wallclock_times }
sys.stdout.write(\
  "%s\n" % json.dumps(js))
```

Listing 2. An elided version of the Python in-process iteration runner (with core-cycles etc. removed).

avoid in-benchmark I/O and memory allocation by storing measurements in a pre-allocated buffer and only writing measurements to `stdout` after all in-process iterations have completed (see Listing 2 for an example). However, the situation on Linux is complicated by our need to read core-cycle and APERF/MPERF counts from Model Specific Register (MSR) file device nodes,[4] which are relatively slow (see Section 7). Since wall-clock time is the most important measure, we ensure that it is the innermost measure taken (i.e. to the extent we control, it does not include the time taken to read core-cycle or APERF/MPERF counts) as shown in Listing 1.

The need to carefully sequence the measurements, and the fact that not all of the VMs in our experiment give us access to the appropriate monotonic timer, meant that we had to implement a small C library (`libkruntime.so`) to do so (see Listing 1 for an example). When possible (all VMs apart from JRuby+Truffle, HHVM, and V8), we used a language's FFI to dynamically load this library; in the remaining cases, we linked the library directly against the VM, which then required us to add user-language visible functions to access them. Core-cycle, APERF, and MPERF counts are 64-bit unsigned integers; since JavaScript and current versions of LuaJIT do not support integers, and since PHP's maximum integer size varies across OS and PHP versions, we convert the 64-bit unsigned measurements to double-precision floating point values in those VMs, throwing an error if this leads to a loss of precision.

## 4 CLASSIFYING WARMUP

The main data created by our experiment is the time taken by each in-process iteration to run. Formally, this is time series data of length 2000. In this Section we explain how we use statistical changepoint analysis to enable us to understand this time series data and classify the results we see, giving the first automated method for identifying warmup.

[4]We forked, and slightly modified, Linux's `msr` device driver to allow us to easily access the MSRs as a non-root user.

### 4.1 Outliers

As is common with analyses of time series data, we first identify *outliers* (in-process iterations with much larger/smaller times than their near neighbours), which in our context are likely to be the result of JIT compilation, GC, or of other processes interrupting benchmarks. Outliers can then be ignored when looking for meaningful performance shifts during changepoint analysis. We use the method described by Tukey [1977], conservatively defining an outlier as one that, within a sliding window of 200 in-process iterations, lies outside the median $\pm 3 \times (90\%\text{ile} - 10\%\text{ile})$. In order that we avoid classifying slow warmup iterations at the start of an execution as outliers (when they are in fact likely to be important warmup data), we ignore the first 200 in-process iterations. Of the 7,320,000 in-process iterations, 0.3% are classified as outliers, with the most for any single process execution being 11.2% of in-process iterations.

### 4.2 Changepoint Analysis

Intuitively, in order to uncover if/when warmup has completed, we need to determine when in-process iteration timings have 'shifted' in nature (i.e. become faster or slower). For example, for traditional warmup, we would expect to see a number of in-process iterations taking time $t$ to be followed by a number at time $t'$ (where $t' < t$). Automating the detection of such 'shifts' is tricky: simple heuristics suffer from both false positives and negatives [Kalibera and Jones 2013]; and 'I know it when I see it' manual detection is time-consuming and prone to human error.

Instead, we use changepoint analysis (see Eckley et al. [2011] for an introduction) to determine if and when warmup has occurred. Formally, a *changepoint* is a point in time where the statistical properties of prior data are different to the statistical properties of subsequent data; the data between two changepoints is a *changepoint segment*. Changepoint analysis is a computationally challenging problem, requiring consideration of large numbers of possible changepoints. A completely naive approach requires an infeasible $2^{n-1}$ calculations ($2^{1999}$ in our context). More sophisticated changepoint analysis algorithms reduce this to $O(n^3)$. We utilise the PELT algorithm [Killick et al. 2012] which reduces the complexity to $O(n)$ by noting that once an 'obvious' changepoint has been discovered, it is not worth including data before that changepoint in further searches.

There are various ways of defining when a changepoint has occurred, but the best fit for our data is to consider changes in both the mean and variance of in-process iterations. To automate this, we use the `cpt.meanvar` function in the R `changepoint` package [Killick and Eckley 2014], passing $15 \log n$ (where $n$ is the time series length minus the number of outliers) to the `penalty` argument, and receiving back changepoint locations along with the mean and variance of each changepoint segment. Whilst dependence is observed in some of our experiment's data, the large penalty we use allows us to make an assumption of independence [Antoch et al. 1997] (see Section 7 for details). Figure 1 shows an example of changepoints and changepoint segments in our context.

### 4.3 Classifications

Building atop changepoint analysis, we can then define useful classifications for time series data from VM benchmarks.

However, we must first acknowledge the constraints that changepoint analysis works under: it has no way of knowing what constitutes the 'noise floor' in our problem domain; nor can it guarantee to find identical means for very similar segments separated from one another by a segment with a clearly distinct mean or variance. To understand how the 'noise floor' affects our analysis, it is helpful to take an extreme example. Consider a long sequence of in-process iterations all timed at 0.1001 seconds, followed by another long sequence timed at 0.1002 seconds. A changepoint will be detected in the transition between the two sequences, resulting in two

```
DELTA = 0.001          # Absolute time delta (in seconds) below which segments are
                       # considered equivalent.
STEADY_STATE_LEN = 500 # How many in-process iterations from the end of the time-series
                       # data will a non-equivalent segment trigger "no steady state"?

def classify(segs):
    assert(len(segs) > 0)
    last_seg = segs[len(segs) - 1]
    lower_bound = last_seg.mean - max(last_seg.variance, DELTA)
    upper_bound = last_seg.mean + max(last_seg.variance, DELTA)
    cls = "flat"
    i = len(segs) - 2
    while i > -1:
        cur_seg = segs[i]
        i -= 1
        if cur_seg.mean + cur_seg.variance >= lower_bound \
          and cur_seg.mean - cur_seg.variance <= upper_bound:
            continue
        elif cur_seg.end > len(segs) - STEADY_STATE_LEN:
            cls = "no steady state"
            break
        elif cur_seg.mean < lower_bound:
            cls = "slowdown"
            break
        assert(cur_seg.mean > upper_bound)
        cls = "warmup"
    return cls
```

Listing 3. The classification algorithm for an individual process execution. Given an ordered list of segments (each with "mean", "variance", and "end" attributes, the latter being the absolute index of the last iteration in the segment) this function returns a string classifying the run sequence's warmup style.

changepoint segments. However, we know that in practice such a small absolute timing delta is as likely to be the result of non-determinism inherent in a real machine or OS as it is to be due to a benchmark or VM: it would thus be better to consider the two segments as being equivalent. Such segments are not always contiguous. Consider a benchmark which, after changepoint analysis, has three segments: the first with a mean of 1.0001 seconds, the second 1.2 seconds (perhaps whilst a JIT compiler is in action), the third 1.0002 seconds. From our perspective, the first and final segments are best considered as being equivalent, even though they are separated by a segment which is clearly different. We therefore need to define a suitable tolerance for determining if two segments are equivalent. In short running benchmarks, a small number of external events can add an absolute level of noise: in our experience readings often vary by a little under 0.001s. In long running benchmarks, an individual external event may add a relatively small amount of noise, but the cumulative effect of many external events can add up. We found that the variance was a good heuristic for this cumulative effect (we simply interpreted it as having units in seconds rather than seconds squared). Combining these two notions, we thus formally define that a segment $s_i$ is equivalent to the final segment $s_f$ if mean($s_i$) is within mean($s_f$) ± max(variance($s_f$), 0.001s).

With that in mind, we can define a simple classification algorithm for a process execution, based on the special nature of the final changepoint segment. Since all our benchmarks run for 2000 in-process iterations we (somewhat arbitrarily) define that a process execution reaches a steady-state if all segments which cover the last 500 in-process iterations are considered equivalent to the final segment. If not, we classify the process execution as *no steady state* (⌇). If a steady state is reached, we are then interested as to whether: all segments are considered equivalent, leading to a classification of *flat* (—); at least one segment is faster than the final segment leading to a classification of *slowdown* (⌡). If a steady state benchmark is not flat or a slowdown, then by definition the final segment must be faster than at least one preceding segment, leading to a classification of *warmup* (⎿). Listing 3 shows an implementation of this algorithm. We consider benchmarks whose behaviour is either flat or warmup as 'good' (flat benchmarks may be unobservably fast warmup), while benchmarks which are either slowdown or no steady state as 'bad'.

Based on this, we can then classify a ⟨VM, benchmark⟩ pair as follows: if its process executions all share the same classification (e.g. warmup) then we classify the pair the same way (in this example, warmup); otherwise we classify the pair as *inconsistent*. There are then two sub-categories of inconsistent benchmarks: 'good' inconsistency (=) is where all a ⟨VM, benchmark⟩ pair's process executions are either flat or warmup; and 'bad' inconsistency (≭) is when one or more process executions are no steady state or slowdown. Good inconsistency can occur because some benchmarks are on the edge of our ability to differentiate warmup from flat behaviour, and we prefer to assume that we are at fault rather than the VM. Bad inconsistency is always more troubling: it means that, at least sometimes, users will experience poor performance.

### 4.4 Steady State Timings

For benchmarks whose process executions all reach a steady state (i.e. are any combination of −, ⅂, or ⅃) we report the number of iterations (*steady iter (#)*) and the time in seconds (*steady iter (s)*) to reach the steady state, as well as the performance within the steady state (*steady perf (s)*).

*Steady iter (#)* and *steady iter (s)* allow one to understand how long it takes a benchmark to reach a steady state, and are made possible by our use of changepoint analysis. Flat process executions by definition have a *steady iter (#)* of 1 and a *steady iter (s)* of 0; we elide these details for benchmarks which are consistently flat. Benchmarks often end up with a variety of distributions for both these measures (often, though not exclusively, seen on benchmarks which contain a mix of flat and non-flat classifications) which makes reporting standard confidence intervals misleading. We therefore use Inter-Quartile Ranges (IQRs) to give an indication of the spread of values, reporting the median and 5% and 95% percentiles (using linear interpolation when the percentile boundaries lie between two data points). Non-overlapping IQRs imply a meaningful difference in the *steady iter (#)* or *(s)* values of a benchmark run on two VMs; however, since IQRs are often widely spread, we are not always able to prove meaningful differences. To give a more nuanced view than IQRs can report we also provide thumbnail histograms.

*Steady perf (s)* roughly corresponds to the 'normal' performance number reported by traditional benchmarking methodologies. In our case, the steady state is composed of one or more segments (see Section 4.3). We report means and 99% confidence intervals calculated via bootstrapping (with 100,000 iterations). Although we assume that values within segments are independent (see Section 7), the values across different segments are clearly not independent. When bootstrapping, we therefore sample values within, but never across, segments (e.g. if a steady state has two segments $A$ and $B$, we bootstrap $A$ to produce $A'$ and $B$ to produce $B'$; we then merge $A'$ and $B'$ to produce a new (bootstrapped) steady state; we never sample values from $A$ into $B'$ or $B$ into $A'$).

## 5 RESULTS

Our results consist of data for 3660 process executions and 7,320,000 in-process iterations. Table 1 summarises the ⟨VM, benchmark⟩ pairs and process executions for each benchmarking machine. Taking $\text{Linux}_{1240\text{v}5}$ as a representative example, 43.4% of ⟨VM, benchmark⟩ pairs have consistently 'good' warmup (i.e. are flat, warmup, or good inconsistent) and 72.4% of all process executions have 'good' warmup (for comparison, $\text{Linux}_{4790}$ is 43.5% and 74.7% respectively and $\text{OpenBSD}_{4790}$ is 30.0% and 86.1% respectively, though the latter runs fewer benchmarks). Fewer ⟨VM, benchmark⟩ pairs than process executions have 'good' warmup because some inconsistent ⟨VM, benchmark⟩ pairs have some 'good' warmup process executions as well as some 'bad'. On $\text{Linux}_{1240\text{v}5}$ 6.5% of ⟨VM, benchmark⟩ pairs slowdown, 6.5% of ⟨VM, benchmark⟩ pairs are no steady state, but the biggest proportion by far is 'bad' inconsistent at 43.5% ($\text{Linux}_{4790}$ has very similar proportions). This latter figure clearly shows a widespread lack of predictability: in almost half of cases, the same benchmark on the same VM on the same machine has more than one performance characteristic. It is tempting to pick one of these performance characteristics – VM benchmarking sometimes reports the fastest process execution, for example – but it is important to note that *all* of these performance characteristics are valid and may be experienced by real-world users.

| Class. | $\text{Linux}_{4790}$ | $\text{Linux}_{1240v5}$ | $\text{OpenBSD}_{4790}$ † |
|---|---|---|---|
| | ⟨VM, benchmark⟩ pairs | | |
| — | 8.7% | 13.0% | 6.7% |
| ˥ | 28.3% | 23.9% | 10.0% |
| ʃ | 6.5% | 6.5% | 0.0% |
| ⁓ | 4.3% | 6.5% | 0.0% |
| = | 6.5% | 6.5% | 13.3% |
| ⋈ | 45.7% | 43.5% | 70.0% |
| | Process executions | | |
| — | 26.4% | 20.9% | 34.0% |
| ˥ | 48.3% | 51.5% | 52.1% |
| ʃ | 16.7% | 17.9% | 11.1% |
| ⁓ | 8.7% | 9.6% | 2.8% |

Table 1. Relative proportions of classifiers across benchmarking machines. Classifiers key: —: flat, ˥: warmup, ʃ: slowdown, ⁓: no steady state, =: good inconsistent, ⋈: bad inconsistent. †Note that $\text{Linux}_{4790}$ and $\text{Linux}_{1240v5}$ run the same set of benchmarks, but $\text{OpenBSD}_{4790}$ runs a subset (see Section 3.3).

Table 1 clearly shows that the results from $\text{Linux}_{1240v5}$ and $\text{Linux}_{4790}$ (both running Linux but on quite different hardware) are comparable, with the relative proportions of all measures being only a few percent different. $\text{OpenBSD}_{4790}$ (running a different OS to $\text{Linux}_{4790}$, but on similar hardware) is however fairly different, with 70.0% of ⟨VM, benchmark⟩ pairs being 'bad' inconsistent. This is partly due to the absence of JRuby+Truffle which, on Linux, is the most consistent (see Section 3.3). Of the benchmarks $\text{OpenBSD}_{4790}$ can run, most behave roughly similarly (as can be seen indirectly from Table 1), but there is a more even spread of process execution classifications across ⟨VM, benchmark⟩ pairs. Tables 8 and 9 in the Appendix contain further details. Overall, we believe that our results are a somewhat clear validation of Hypothesis H2.

Looking at one machine's data brings out further detail. For example, Table 2 (with data from $\text{Linux}_{1240v5}$) enables us to make several observations. First, of the 6 benchmarks, only *n-body* and *spectralnorm* come close to 'good' warmup behaviour on all VMs (though each has 1 or 2 bad inconsistent cases). Second, *binary trees* seems generally to be bad inconsistent, and *fasta* and *Richards* often bad inconsistent. This raises the question as to whether some aspect of these benchmarks makes inconsistent performance more likely, though they share no particularly obvious characteristics. Third, the overall spread of steady state iter (#) values is uneven: VMs seem to mostly either reach a steady state quickly (often in 10 or fewer in-process iterations) or take hundreds of in-process iterations. The latter are troubling because previous benchmarking methodologies will not have run the benchmarks long enough to see the steady state emerge. Fourth, it does not seem to be the case that VMs which take a long time to reach a steady state have faster steady states: if anything, the opposite appears to be true, suggesting that users of some code on some VMs suffer from a 'double whammy' of poor warmup and poor steady state performance.

| |  | Class. | Steady iter (#) | Steady iter (s) | Steady perf (s) |  | Class. | Steady iter (#) | Steady iter (s) | Steady perf (s) |
|---|---|---|---|---|---|---|---|---|---|---|
| C | *binary trees* | ⏦ | | | | *n-body* | — | | | 0.40555 ±0.005110 |
| Graal | | ≍ (27⅂, 3ʃ) | 32.0 (17.0, 193.8) | 6.60 (3.729, 36.608) | 0.18594 ±0.000315 | | ⅂ | 8.0 (7.5, 8.5) | 1.22 (1.126, 1.358) | 0.13334 ±0.000045 |
| HHVM | | ≍ (24⅂, 4ʃ, 2⏦) | | | | | ≍ (16⅂, 11ʃ, 3⏦) | | | |
| HotSpot | | ≍ (25⅂, 5ʃ) | 7.0 (7.0, 53.5) | 1.19 (1.182, 9.703) | 0.18279 ±0.000116 | | ⅂ | 2.0 (2.0, 2.0) | 0.14 (0.141, 0.143) | 0.13699 ±0.000032 |
| JRuby+Truffle | | ʃ | 1082.0 (999.0, 1232.5) | 2219.59 (2039.304, 2516.021) | 2.05150 ±0.017738 | | ⅂ | 69.0 (69.0, 70.0) | 17.95 (17.716, 18.127) | 0.20644 ±0.001568 |
| LuaJIT | | ≍ (23⅂, 4ʃ, 2—, 1⏦) | | | | | — | | | 0.25399 ±0.004471 |
| PyPy | | ≍ (27ʃ, 3⏦) | | | | | — | | | 1.85835 ±0.012893 |
| V8 | | ≍ (15—, 9⅂, 6ʃ) | 1.5 (1.0, 794.0) | 0.25 (0.000, 391.026) | 0.49237 ±0.003198 | | = (25—, 5⅂) | 1.0 (1.0, 361.6) | 0.00 (0.000, 87.367) | 0.24138 ±0.000389 |
| C | *fannkuch redux* | ≍ (21—, 6⅂, 2ʃ, 1⏦) | | | | *Richards* | ≍ (19—, 5ʃ, 4⏦, 2⅂) | | | |
| Graal | | ≍ (28⅂, 1⏦, 1ʃ) | | | | | ≍ (28⅂, 1⏦, 1ʃ) | | | |
| HHVM | | ⅂ | 10.0 (10.0, 10.0) | 52.66 (52.660, 52.708) | 1.35779 ±0.011948 | | | | | |
| HotSpot | | ⅂ | 390.0 (2.0, 390.0) | 153.70 (0.407, 155.254) | 0.36202 ±0.022767 | | ≍ (26ʃ, 2⏦, 2⅂) | | | |
| JRuby+Truffle | | ʃ | 1016.5 (999.0, 1023.1) | 1039.04 (1014.290, 1059.967) | 1.08833 ±0.038580 | | ʃ | 1021.0 (1014.9, 1027.0) | 917.30 (901.708, 946.683) | 0.89509 ±0.027590 |
| LuaJIT | | — | | | 0.56285 ±0.000097 | | ≍ (21⏦, 7ʃ, 2⅂) | | | |
| PyPy | | ≍ (15⅂, 13—, 2ʃ) | 2.0 (1.0, 28.9) | 1.57 (0.000, 43.483) | 1.55442 ±0.020549 | | ⅂ | 2.0 (2.0, 5.0) | 1.12 (1.114, 4.054) | 0.96809 ±0.010950 |
| V8 | | = (19⅂, 11—) | 2.0 (1.0, 25.5) | 0.31 (0.000, 7.525) | 0.30401 ±0.000154 | | = (29⅂, 1—) | 4.0 (4.0, 16.0) | 1.44 (1.434, 7.135) | 0.47421 ±0.001218 |
| C | *fasta* | — | | | 0.07048 ±0.000210 | *spectralnorm* | ≍ (28⅂, 1—, 1ʃ) | 997.0 (2.0, 1001.0) | 546.38 (0.546, 547.717) | 0.54547 ±0.002562 |
| Graal | | ≍ (29⅂, 1⏦) | | | | | ≍ (29ʃ, 1—) | 15.0 (2.0, 21.0) | 12.40 (0.812, 17.804) | 0.89293 ±0.000087 |
| HHVM | | ≍ (27⅂, 2⏦, 1ʃ) | | | | | ⅂ | 35.0 (34.0, 41.0) | 139.18 (136.909, 147.626) | 1.40690 ±0.011133 |
| HotSpot | | ≍ (18⅂, 12ʃ) | 261.0 (6.0, 595.0) | 30.73 (0.614, 70.021) | 0.11744 ±0.001723 | | ⅂ | 7.0 (7.0, 8.6) | 1.90 (1.901, 2.425) | 0.31470 ±0.000029 |
| JRuby+Truffle | | | | | | | ⅂ | 1011.0 (1007.5, 1014.0) | 893.38 (888.985, 896.562) | 0.83633 ±0.014403 |
| LuaJIT | | ⏦ | | | | | — | | | 0.22435 ±0.000020 |
| PyPy | | ⏦ | | | | | ⅂ | 75.0 (75.0, 75.0) | 34.43 (34.429, 34.437) | 0.46489 ±0.000046 |
| V8 | | ≍ (19⅂, 10ʃ, 1⏦) | | | | | ⅂ | 3.0 (3.0, 3.0) | 0.55 (0.554, 0.554) | 0.24963 ±0.000039 |

Table 2. Results for $\text{Linux}_{1240v5}$. We report the classification of each process execution and, for inconsistent benchmarks, the constituent classifications (e.g. ≍ (22ʃ, 8⅂) means 22 slowdowns and 8 warmups). If a steady state is achieved, we report three summary statistics. *Steady iter (#)* is the median in-process iteration number to reach the steady state, and *Steady iter (s)* is the median wall-clock time (since the beginning of the process execution) to reach the steady state; for both measures we report 5% and 95% inter-quartile ranges. *Steady perf (s)* is the mean steady state performance across all process executions (reported with 99% confidence intervals). Thumbnail histograms show the spread of values; the red bars indicates the median values.

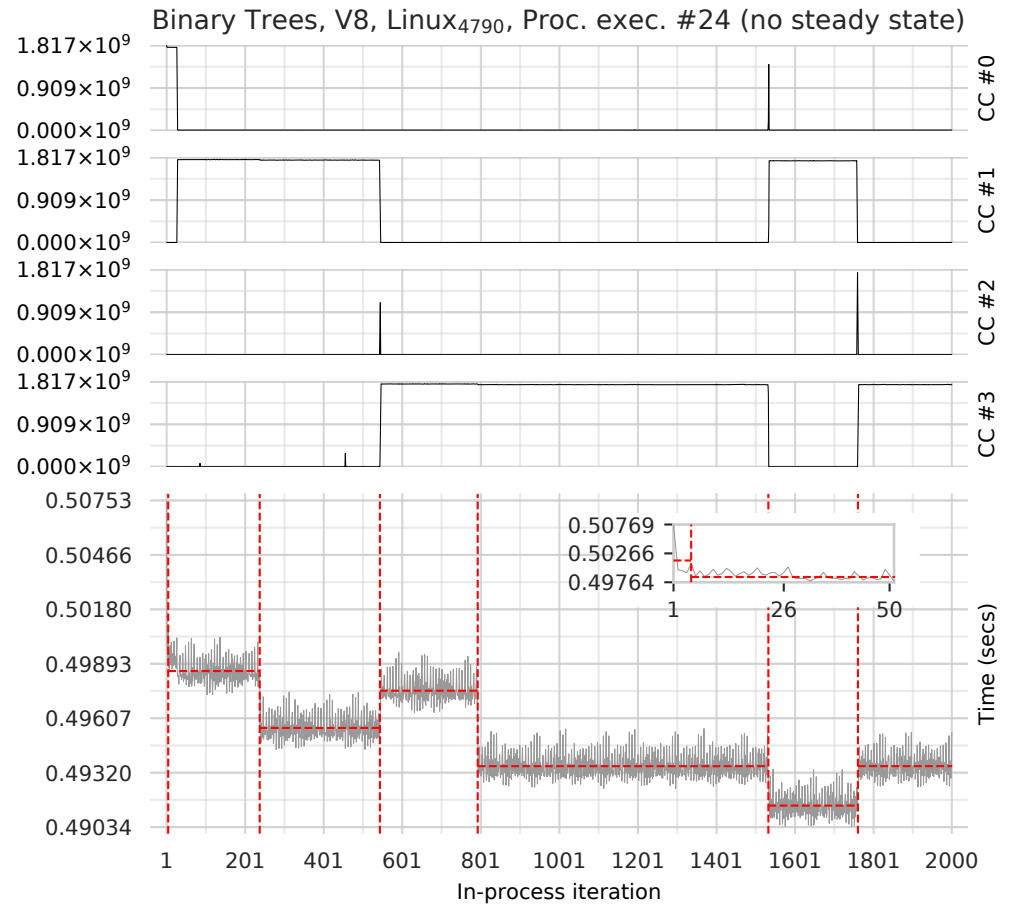

Fig. 2. An example of a run-sequence plot with core-cycle plots (one for each of the 4 CPU cores; note that they all have the same $y$ axis scale). In this example we can clearly see a benchmark migrating between cores. In some cases this migration also aligns with a clear changepoint, though not in others. The continual changepoints lead to an overall classification of no steady state.

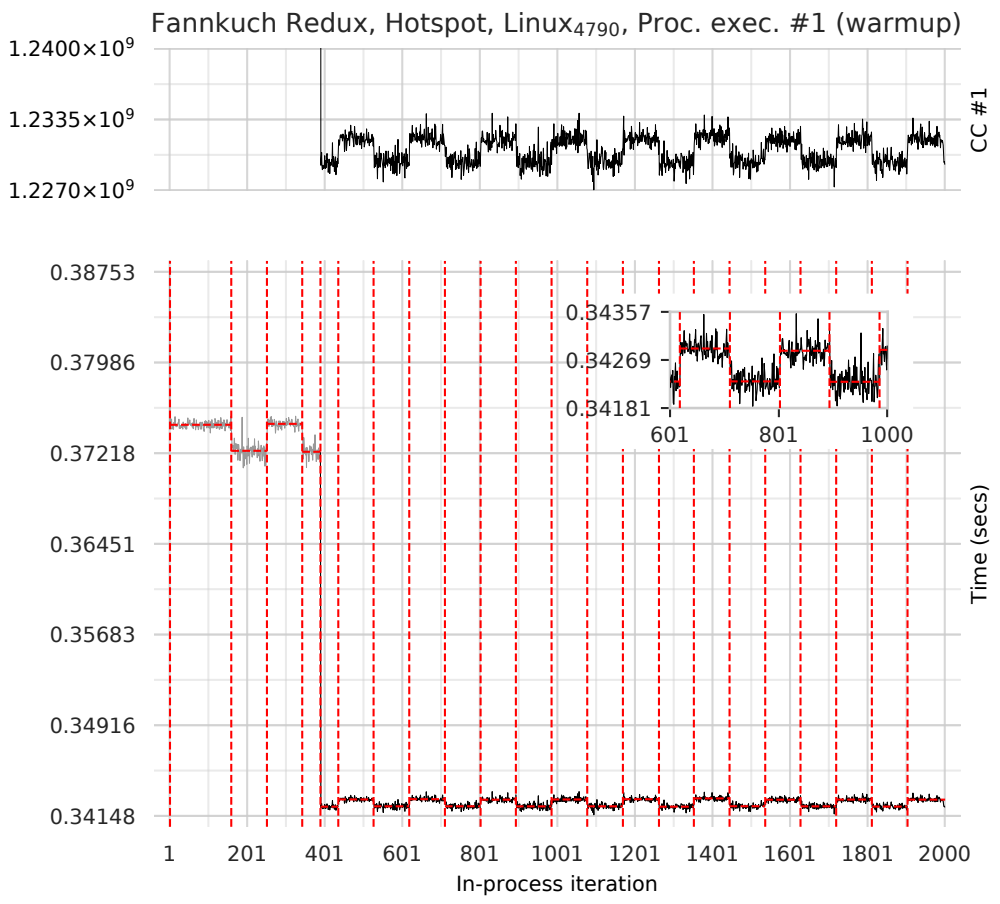

Fig. 3. Cycles in wall-clock times which are reflected by the core-cycle count for the relevant core. Although the changepoint analysis has captured the cycles, the change in performance is too small for the classification algorithm to consider this example 'no steady state'.

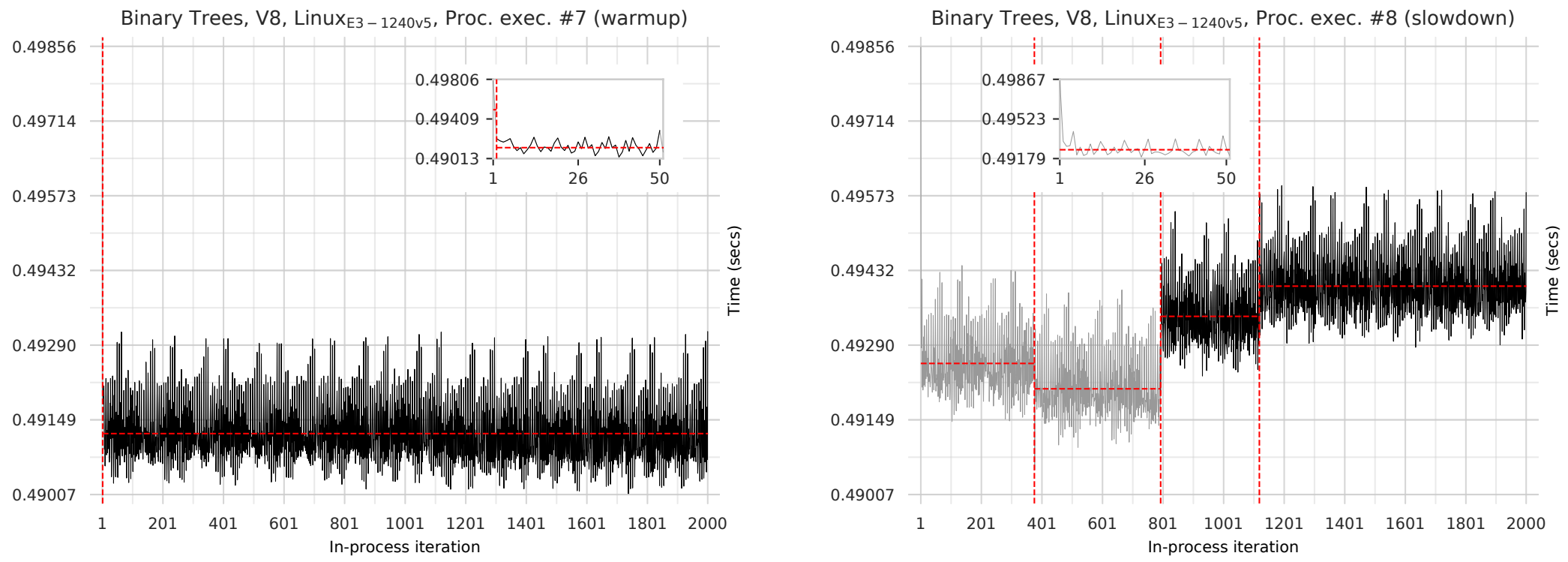

Fig. 4. An example of bad inconsistency: the same benchmark, on the same VM, on the same machine, with one process execution warming up and the other slowing down.

### 5.1 Warmup Plots

We created several different types of plot to help us understand our data in detail. All plots have in-process iteration number on the $x$-axis. Run-sequence plots show wall-clock times on the $y$-axis. The plots in Figure 1 show run-sequence plots for warmup and slowdown behaviours. Figure 4 shows an example of bad inconsistency and Figure 5 of good inconsistency. Core-cycle plots (Linux only) show the core-cycle counts on the $y$-axes (one plot per-core). The plots in Figures 2 and 3 show examples of run-sequence plots accompanied by core-cycle plots for no steady state and cyclic data respectively.

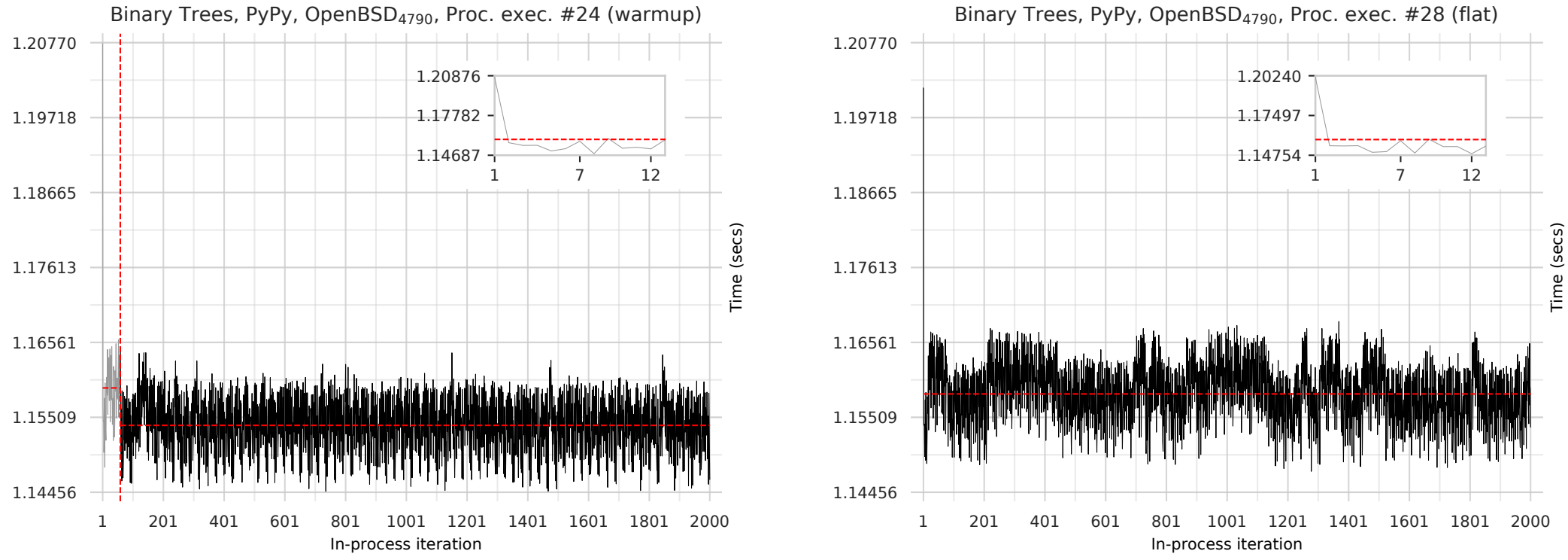

Fig. 5. An example of good inconsistency where one process execution is classified as warmup, the other flat. Although a human might consider both of these as warmup, there is a subtle difference in the nature of the first 20–30 in-process iterations in the LHS and RHS plots, which leads changepoint analysis to consider the RHS plot as a single changepoint segment.

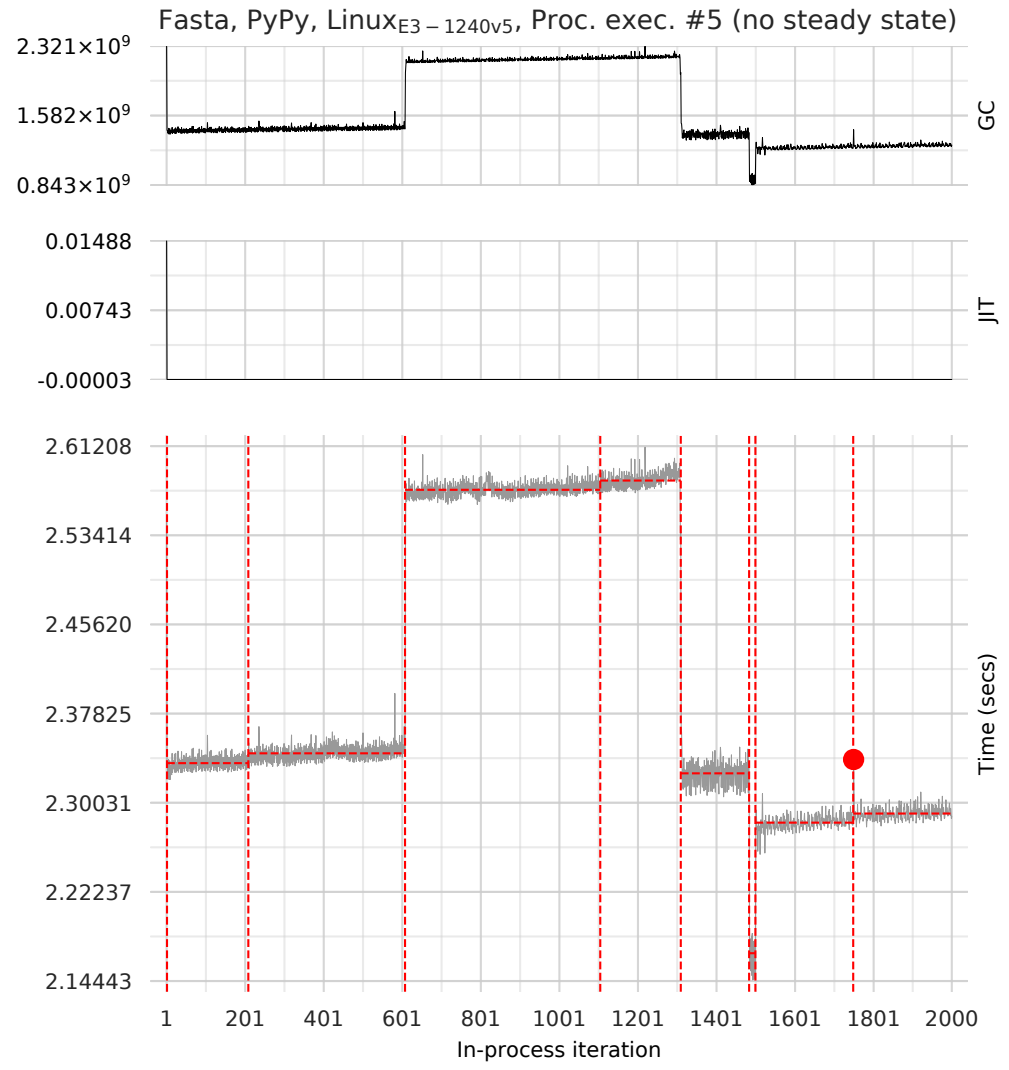

Fig. 6. A run-sequence plot (bottom) with GC (top) and JIT (middle) plots. In this case, the benchmark's gradually increasing wall-clock time clearly correlates with GC. This example (along with several others) appears to show a memory leak in PyPy, which we have reported upstream.

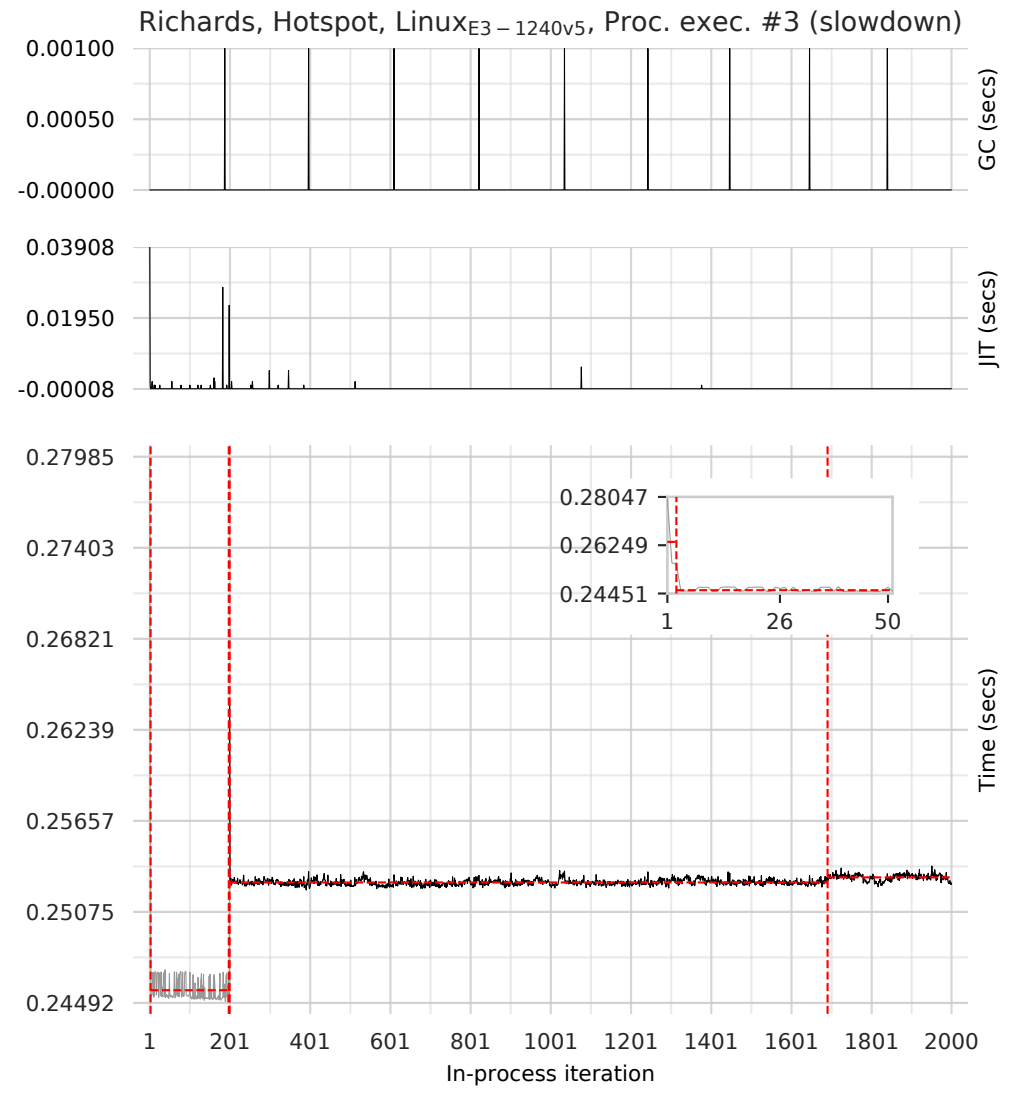

Fig. 7. Slowdown at in-process iteration #199, which correlates with two immediately preceding JIT compilation events that may explain the drop in performance. Note that the regular GC spikes have only a small effect on wall-clock time.

Core-cycle plots help us understand how VMs use, and how the OS schedules, threads. Benchmarks running on single-threaded VMs are characterised by a high cycle-count on one core, and very low (though never quite zero) values on all other cores. VMs may migrate between cores during a process execution, as can be clearly seen in Figure 2. Although multi-threaded VMs can run JIT compilation and / or GC in parallel, it is typically hard to visually detect such parallelism as it tends to be accompanied by frequent migration between cores. However, it can often be seen that several cores are active during the first few in-process iterations whilst JIT compilation occurs.

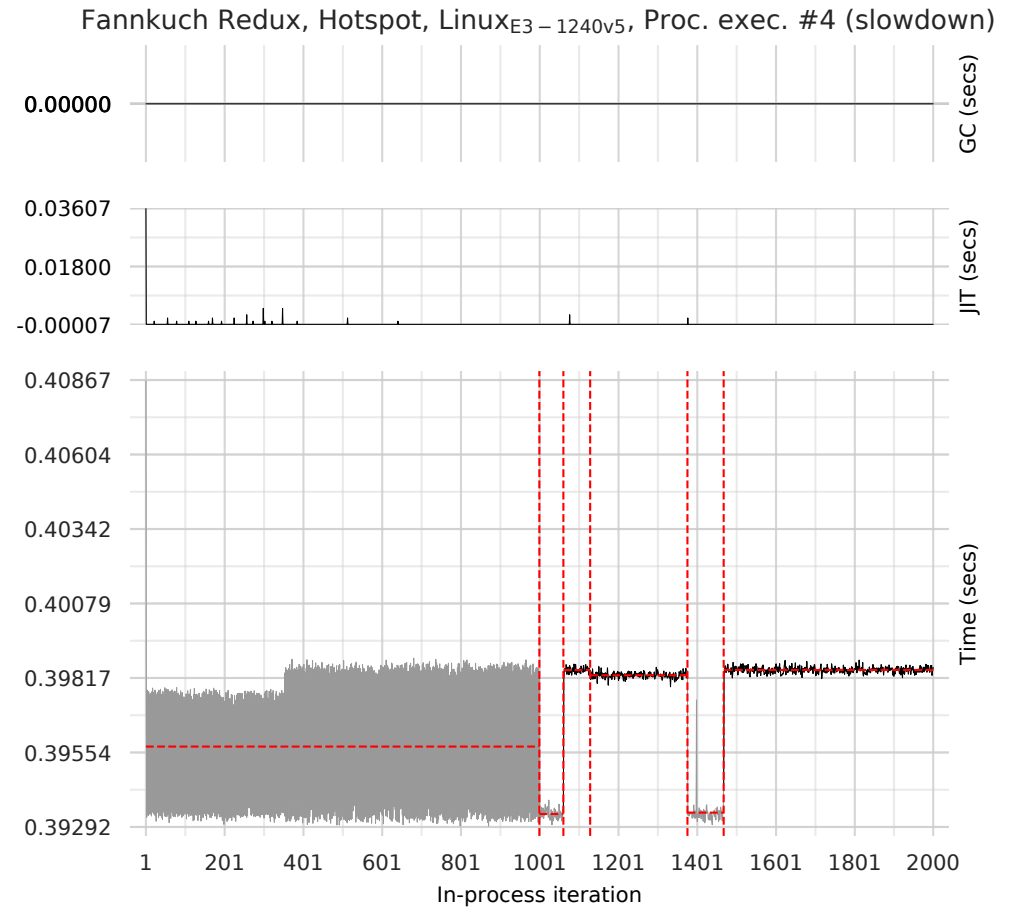

Fig. 8. An example where performance changes do not appear to correlate with compilation or GC.

| VM | Machine | |
|---|---|---|
| | Linux$_{4790}$ | OpenBSD$_{4790}$ |
| C | 0.00075 ±0.000029 | 0.00091 ±0.000002 |
| Graal | 0.34815 ±0.000233 | |
| HHVM | 0.75270 ±0.002056 | |
| Hotspot | 0.14716 ±0.000708 | 0.17997 ±0.000247 |
| JRubyTruffle | 2.66179 ±0.011864 | |
| LuaJIT | 0.00389 ±0.000442 | 0.00429 ±0.000003 |
| PyPy | 0.44040 ±0.001067 | 0.78696 ±0.000481 |
| V8 | 0.08727 ±0.000239 | 0.12016 ±0.000084 |

Table 3. VM startup time (in seconds with 99% confidence intervals).

### 5.2 The Effects of Compilation and GC

The large number of non-warmup cases in our data led us to make the following hypothesis:

**H3** Non-warmup process executions are largely due to JIT compilation or GC events.

To test this hypothesis, we made use of the debug facilities of HotSpot and PyPy to record the time spent performing JIT compilation and GC. Since recording this additional data could potentially change the results we collect, it is only collected when Krun is explicitly set to 'instrumentation mode'. This allows us to identify interesting correlations (though we make no claim that they prove causation). For example, Figure 6 shows an example where slowdown is clearly correlated to garbage collection. Similarly, in Figure 7, there is a clear correlation between a JIT compilation and a slowdown. However, in many cases such as Figure 8 neither JIT compilation, nor garbage collection, are sufficient to explain odd behaviours.

The relatively few results we have with GC and JIT compilation events, and the lack of a clear message from them, means that we feel unable to validate or invalidate Hypothesis H3. Whilst some non-warmups are plausibly explained by GC or JIT compilation events, many are not, at least on HotSpot and PyPy. When there is no clear correlation, we have very little idea of a likely cause of the unexpected behaviour. It may be that obtaining similar data from other VMs will clarify this issue, but not all VMs support this feature and, in our experience, those that do support it do not always document it accurately.

## 6 STARTUP TIME

The data presented thus far in the paper has all been collected after the VM has started executing the user program. The period between a VM being invoked and it executing the first line of the user program is the VM's *startup* time, and is an important part of a VM's real-world performance.

A small modification to Krun enables us to measure startup. We prepend each VM execution command with a small C wrapper, which prints out wall-clock time before immediately executing the VM itself; and, for each language under investigation, we provide a 'dummy' iterations runner which simply prints out wall-clock time. In other words, we measure the time just before the VM is loaded and at the first point that a user-level program can execute code on the VM; the delta between the two is the startup time. For each VM we run 200 process executions (for startup, in-process iterations are irrelevant, as the user-level program completes as soon as it has printed out wall-clock time).

Table 3 summarises our startup results. Since Krun reboots before each process execution, these are measures of a 'cold' start and thus partly reflect disk speed etc. (though our machines load VMs from an SSD). As this data clearly shows, startup time varies significantly amongst VMs: taking C on Linux as a baseline, the fastest VM (LuaJIT) is 5× slower, whilst the slowest VM (JRuby+Truffle) is around 3500× slower, to startup.

## 7 THREATS TO VALIDITY

While we have designed our experiment as carefully as possible, we do not pretend to have controlled every possibly confounding variable. It is inevitable that there are further confounding variables that we are not aware of, which may have coloured our results.

We have tried to gain an understanding of the effects of different hardware on benchmarks by using machines with the same OS but different hardware. More distinct hardware (e.g. a non-x86 architecture) is likely to uncover further hardware-related differences. However, hardware cannot be varied in isolation from software: the greater the differences in hardware, the more likely that JIT compilers are to use different components (e.g. different code generators). Put another way, an apples-to-apples comparison across very different hardware is often impossible, because the 'same' software is itself different.

We have not systematically tested whether rebuilding VMs affects warmup, an effect noted by Kalibera and Jones, though which seems to have little effect on the performance of JIT compiled code [Barrett et al. 2015]. However, since measuring warmup largely involves measuring code that was not created by a JIT compiler, it is possible that these effects may affect our experiment. To a limited extent, the rebuilding of VMs that occurred on each of our benchmarking machines gives some small evidence as to this effect, or lack thereof.

The checksums we added to benchmarks ensure that, at a user-visible level, each benchmark performs equivalent work in each language variant. However, it is impossible to say whether each performs equivalent work at the lowest level or not. For example, choosing to use a different data type in a language's core library may substantially impact performance. There is also the perennial problem as to the degree to which an implementation of a benchmark should respect other language's implementations or be idiomatic (the latter being likely to run faster). From our perspective, this is somewhat less important, since we are interested in the warmup patterns of reasonable programs, whether they are the fastest possible or not. It is however possible that by inserting checksums we have created unrepresentative benchmarks, though this complaint could arguably be directed at the unmodified benchmarks too.

Although we have minimised the number of system calls that our in-process iterations runners make, we cannot escape them entirely. For example, on both Linux and OpenBSD `clock_gettime` (which we use to obtain monotonic wall-clock time) contains what is effectively a spin-lock, meaning that there is no guarantee that it returns within a fixed bound. Fortunately, in practise, `clock_gettime` returns far quicker than the granularity of any of our benchmarks. The situation on Linux is complicated by our reading of core-cycle, APERF, and MPERF counters via MSR device nodes: we call `lseek` and `read` between each in-process iteration (we keep the files open across all in-process iterations to reduce the open/close overhead). These calls are more involved than `clock_gettime`: as well as the file system overhead, reading from a MSR device node triggers inter-processor interrupts to schedule an `RDMSR` instruction on the desired core (causing the running core to save its registers etc.). We are not aware of a practical way to lower these costs.

Although Krun does as much to control CPU clock speed as possible, modern CPUs do not always respect operating system requests. On Linux, we use the APERF/MPERF ratio to check for frequency changes. Although the hoped-for ratio is precisely 1, there is often noticeable variation around this value due to the cost of reading these counters. On active cores the error is small (around 1% in our experience), but on idle cores (which are rarely truly idle, as occasional computation happens on them) it can be relatively high (10% or more is not uncommon; in artificially extreme cases, we have observed up to 200%). We first need to exclude idle cores from any checks. Formally the rate at which APERF increments is undefined [Intel 2017]. In practise, however, our machines increment APERF on a fully utilised core at the CPU base frequency (e.g. $\text{Linux}_{4790}$'s CPU increments APERF at 3.6GHz). We therefore define an idle core to be one whose APERF is 1000× less than this (per second). We then need to determine a sensible tolerance around the ideal APERF/MPERF ratio of 1, taking into account the error introduced by reading these counters. For the 4790 processor, the frequency values either side of its 3.6GHz base are 3.4GHz and 3.8GHz, which would lead to a 5% difference in the APERF/MPERF ratio (for the 3.5GHz E3-1240 processor, the values either side are 3.3GHz and 3.9GHz i.e. at least a 6% difference in the APERF/MPERF ratio). Based on this, we define the safe APERF / MPERF ratio to be $1 \pm 0.03$.[5] We then wrote a simple tool to examine every process execution on $\text{Linux}_{4790}$ and $\text{Linux}_{1240v5}$, checking that every active core had a safe APERF/MPERF ratio. Of the 5,520,000 in-process iterations on these two machines, 225 (0.004%) spread over 10 process executions failed this check, all but 2 on $\text{Linux}_{1240v5}$. Fortunately, outlier detection is designed to deal with such aberrations, and 216 (98%) of these in-process iterations are detected as outliers. Since outliers do not affect changepoint analysis or the statistics we produced, 8 of the 10 process executions are not affected in practise. Since the 2 in-process iterations on $\text{Linux}_{4790}$ do not have noticeably different wall-clock times than their neighbours, and both have an APERF/MPERF ratio of 1.031 (only marginally above our safe threshold), we do not consider them problematic. The remaining 7 in-process iterations which are not detected as outliers are contained in 1 process execution of Graal running *fasta*: manual inspection shows that these in-process iterations are fairly evenly spread over the (somewhat noisy) process execution and that two of them led to a small changepoint segment that causes the process execution to be classified as no steady state rather than warmup. As this ⟨VM, benchmark⟩ pair was ⋊ (29⌐, 1≁), it is likely that if the processor had run at full speed, the overall benchmark would have been classified as warmup (as indeed it was on $\text{Linux}_{4790}$). However, overall, the similar figures for $\text{Linux}_{4790}$ and $\text{Linux}_{1240v5}$ shown in Table 1 are strong evidence of the limited effect this is likely to have had on our results.

Our experiments allow the kernel to run on the same core as benchmarking code. We experimented extensively with CPU pinning, but eventually abandoned it. After being confused by the behaviour of Linux's `isolcpus` mechanism (whose semantics changes between a normal and a real-time kernel), we used CPU shielding (`cset shield`) to pin the benchmarks to the 3 non-boot cores on our machines. However, we then observed notably worse performance for VMs such as HotSpot. We suspect this is because such VMs query the OS for the number of available cores and create a matching number of compilation threads; by reducing the number of available cores, we accidentally forced two of these threads to compete for a single core's resources.

Address Space Layout Randomisation (ASLR) is, from a performance perspective, a controversial feature. It clearly introduces significant non-determinism into benchmarking, and it is thus tempting

[5]This means that small portions of an in-process iteration could be subject to frequency changes yet still fall within this tolerance. Given the error in reading the APERF and MPERF counters, this is an inevitable problem; by making our tolerance relatively tight, we minimise the chances of missing genuine issues.

to turn it off. However, this then raises the very real possibility that the one point in the ASLR space one is sampling from is not indicative of the entire space. Ideally, we would use a system like Stabilizer [Curtsinger and Berger 2013] to sample across the search space evenly; unfortunately, we were not able to build Stabilizer on a modern Linux system.[6] Since ASLR cannot be turned off on OpenBSD, we left ASLR on Linux on, to make a comparison across the two easier. By measuring 30 process executions, we lessened, though did not remove, the chances of all our measurements coming from a narrow part of the overall search space.

In controlling confounding variables, our benchmarking environment inevitably deviates from standard configurations. It is possible that in so doing we have created a system where benchmarks warmup (or otherwise) in a way that few will ever see in practise. However, our judgement is that this is a reasonable trade-off, given that the alternative of a system that is likely to introduce substantial noise into our readings.

Current changepoint analysis software available typically make an assumption of independence within and across changepoint segments. The `cpt.meanvar` function we use also makes this assumption of independence, which our in-process iteration timings are clearly not. We can safely use a changepoint approach that assumes independence by using a 'larger' penalty value [Antoch et al. 1997]. As this suggests, penalties are as much an art as a science. A typical penalty for our setup is $4 \log n$ whereas we used $15 \log n$. We manually inspected a large number of benchmarks and found that the changepoints identified were rarely much worse, and in the vast majority of cases better than, our initial manual attempts. Given the large number of plots to analyse (3660 for the main part of the experiment), it is extremely unlikely that we would have been able to do anywhere near as good a job manually as changepoint analysis does automatically.

Although we only ever bootstrap within, and never across, segments (see Section 4.4), we are forced to assume that data within segments are independent. Across all three of our benchmarking machines, 11.8% of process executions showed some form of dependence in one or more changepoint segments. The consequences of this are limited for two reasons. First, over 99% of steady state segments have a variance under 0.001s (with the vast majority one or more orders of magnitude smaller than that), which means that the confidence intervals we produce will be small no matter what we do. Second, for our style of data the confidence intervals we produce when dependence is present tend to demonstrate a small over-approximation. To understand this, we ran a simulation study with the smallest (-0.968), largest (0.668), and mean (-0.215) dependence we observed (where 0 is independence and $\pm 1$ is lag 1 correlation). The coverage of the 99% confidence intervals over 1,000 simulations was 100%, 92%, and 99.9% respectively (where values over 99% represent negative correlation and values under 99% represent positive correlation). As a comparison, we also simulated independent data (using the same seed as used in the dependent simulations) which had coverage of 98.3%, showing that even with independent data one may not always get precisely the expected coverage. Overall, despite these limitations, we believe that our approach produces statistics that are more representative of a benchmark's performance than current approaches.

## 8 RELATED WORK

There are two works we are aware of which explicitly note unusual warmup patterns. Whilst running benchmarks on HotSpot, Gil et al. [2011] observed inconsistent process executions (e.g. recursiveErgodic), and benchmarks that we could classify as no steady state (listBubbleSort) and slowdown (arrayBubbleSort). By running a greater number of (somewhat larger) benchmarks on a number of VMs, and executing them in a more tightly controlled execution environment, our results can be seen as significantly strengthening Gil et al.'s observations. Our work also adds an automated approach to identifying when warmup has occurred and classifying run-sequence plots.

[6]Stabilizer's authors hope to produce a new version which is more easily ported across OS versions, though with no definite timescale.

[Kalibera and Jones 2013] note the existence of what we have called cyclic behaviour, which causes them to classify a benchmark as not reaching an independent state and thus being unsuitable for bootstrapping as-is. In such instances, they require the user to manually pick one part of the cycle (i.e. always pick iteration *n* after the steady state of each process execution) for bootstrapping, which leads to a clear potential for bias (for one benchmark, one could pick the slow part of the cycle, for another the fast part). As our approach automatically identifies changepoint segments, we take a very different approach. Changepoint analysis naturally identifies 'large' cycles (i.e. where the cycles are fairly lengthy and/or with substantial differences between the means of the top and bottom of the cycle) as different segments with non-equivalent means, which then leads to a classification of no steady state. 'Small' cycles either lead to different segments with equivalent means or, more often, to a single segment.

Probably the most widely used method for detecting a steady state is that of Georges et al. [2007] which looks for in-process iterations where the coefficient of variance (standard deviation divided by the mean of the relevant measurements) falls below a threshold. Using a threshold of 0.01 – the more conservative of the two values suggested by Georges et al. – we applied their heuristic to our data. In virtually all of the cases where we found a steady state, Georges et al.'s heuristic also finds a steady state (though not necessarily at the same point). However, it also finds steady states for 78.1% of the process executions we classify as no steady state, including Figure 2. This confirms the findings from [Kalibera and Jones 2013] that simple heuristics can often give misleading statistics for VM benchmarking.

## 9 SUGGESTIONS FOR THE COMMUNITY

We do not claim that this paper represents the end of the road in terms of VM benchmarking methodologies: we hope and expect that this work will be superseded by better approaches in the future. However, imperfect though it is, this paper does clearly show problems with previous approaches to VM benchmarking, including those this paper's authors have been involved with in the past. In this section we make some tentative suggestions to the VM developer and user communities based on our experience.

First, our results undermine the previous VM benchmarking orthodoxy of benchmarks quickly and consistently reaching a steady state after a fixed number of iterations. It is clear that many benchmarks take considerable time to reach a steady state; that different process executions of the same benchmark reach a steady state at different points; and that some process executions do not ever reach a steady state. Even if one were to pick a very high number of initial in-process iterations to discard, there is no guarantee that the remainder would represent the steady state. Similarly, our results also show that one cannot assume that one process execution of a ⟨VM, benchmark⟩ pair is representative of others: each process execution must be analysed individually to determine if, and when, a steady state is reached. We believe that, in all practical cases, this means that one must use an automated approach to analysing each process execution individually. The open-source changepoint analysis approach presented in this paper is one such option.

Second, we believe that the traditional practise of presenting only steady state numbers is hard to defend. There are cases in our results where, for a given benchmark, two or more VMs have steady state performance within 2x of each other, but warmup differs by 100-1000x. While some users may wish to trade warm-up time for better steady state performance, others may not. Furthermore, there does not seem to be a clear relationship between these two factors: VMs which are slower to warmup often have worse steady state performance! We thus hope that future VM experiments will report: a benchmark's classification (warmup, flat, slowdown, no steady state); and, if it reaches a steady state, when the steady state was reached (using one or both of steady state iter (# or s)[7]) and the performance of the steady state (steady perf (s)).

Third, it is important to make benchmarks run long enough that any measurements taken are substantially above the noise floor. It is not uncommon to see published benchmarks that run for 0.001s or smaller: external events such as a context switch can have a significant (relative) effect on such small measurements, giving one a false impression of performance. Because of this, in our experiment changepoint segments with means less than 0.001s apart are considered equivalent. However, running in-process iterations for long periods of time is also somewhat undesirable. When possible, we suggest that the fastest in-process iterations should run for around 0.5s. However, this can sometimes be impractical: when benchmarking a set of VMs, the slowest VM may be an order of magnitude slower than another; even within a single VM, in-process iterations before warmup can be one or two orders of magnitude slower than after warmup. In such cases we suggest the minimum acceptable time for an in-process iteration is 0.1s.

Fourth, we suspect that some of the odd results we have seen result from over-training VM heuristics on small sets of benchmarks. The machine-learning community's approach to the problem of over-training may apply equally well to VMs: this would involve using a training set of benchmarks to devise heuristics, and then benchmarking the resulting system(s) on a separate validation set of benchmarks.

Fifth, we suspect that a reliance on small suites of benchmarks means that only small parts of VMs are being benchmarked effectively. We are increasingly of the opinion that benchmarking quality and quantity are tightly related, and that VMs need to be run on many more benchmarks than is typically the case. However, collecting benchmarks is surprisingly difficult as there are few guidelines as to what makes a good or bad benchmark, and the more constraints one places on 'acceptable' benchmarks, the harder the process becomes. For example, in this paper we ensured that our benchmarks were deterministic. In modern languages, where semi-random hashing of data-structures is common, many moderately sized candidate benchmarks will be non-deterministic. As benchmark suites grow in size, such trade-offs are likely to become more frequent: one pragmatic solution may simply be to note which benchmarks have such properties and to allow readers to take that into account when evaluating performance.

## 9.1 How Long Should Experiments Be Run For?

Of all the techniques we have used in this paper, the most simple was the most important: we simply ran benchmarks for much longer than suggested by previous methodologies. Traditional VM experiments often run for 5 or 10 in-process iterations. Our results clearly show that, when a statistically robust analysis is desired, such short runs are insufficient. However, none of us wants to run benchmarks for any longer than necessary.

In this subsection we sample data from $\text{Linux}_{1240v5}$ to see if fewer process executions or in-process iterations give similar results to the full results. Although the resulting analysis necessarily only applies to our experiment, we are able to give some tentative pointers to those benchmarking other systems, though we strongly caution against assuming these will hold in every circumstance.

We define 'similar' as "what percentage of the sampled ⟨VM, benchmark⟩ pairs are statistically equivalent to those in Table 2?" We report both the individual characteristics (classification, steady iter (# or s), and steady perf (s)) and an 'overall' comparison, which records the percentage of ⟨VM, benchmark⟩ pairs where all three individual characteristics are statistically equivalent. Checking for equivalence amounts to checking whether confidence intervals overlap.[8] For classifications, we compute multinomial proportion confidence intervals (i.e. the proportion of warmup, slowdown, flat and no steady state classifications) using the method of [Sison and Glaz 1995].

[7] Since steady iter (#) and steady iter (s) report the same underlying measure in different units they can be treated as one.
[8] Note that there is an unavoidable source of error in any comparison involving confidence intervals since there is no guarantee that the original confidence interval covers the 'true' mean. Thus sometimes the sampled ⟨VM, benchmark⟩ pair

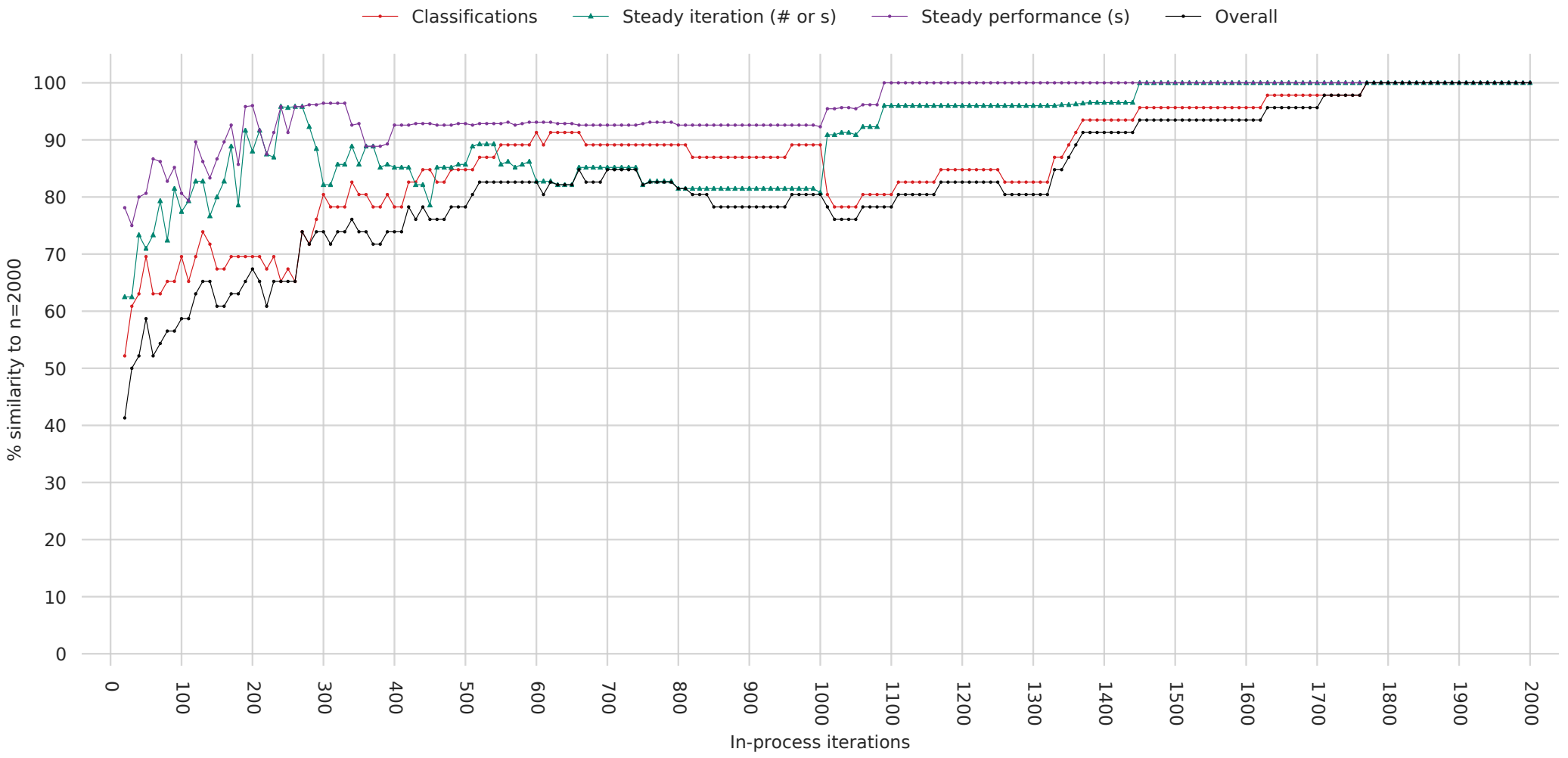

Fig. 9. How similar $\text{Linux}_{1240v5}$'s results (seen in Figure 2) would have been if we had run fewer than 2000 in-process iterations. We truncated process executions (taking in-process iterations $0 \ldots n$ where $n \in \{10, 20, \ldots, 1980, 1990\}$) and compared their classifications, time to reach a steady state, and steady state performance to the original ($n = 2000$). The 'overall' similarity of each truncated dataset is the number of ⟨VM, benchmark⟩ pairs for which all three characteristics are statistically equivalent to the full experiment.

While in most cases the comparisons involved are straightforward, no steady state process executions are trickier. Consider a ⟨VM, benchmark⟩ pair $P$ from the original results and a ⟨VM, benchmark⟩ pair $P_s$ sampled from $P$. What should we do if $P$ contains a no steady state process execution but $P_s$ does not (or vice versa)? In such cases, the classifications comparison is made as normal (using the confidence intervals) but we do not count the steady iter (# or s) and steady perf (s) comparisons as either succeeding or failing, since no meaningful comparison can be made.

*9.1.1 In-process iterations.* We chose 2000 in-process iterations based on our experience of VM internals. VMs have many counters which, when they exceed a certain threshold, cause the VM to change behaviour (e.g. (re)compile code, change garbage collection strategy etc.). Our experience is that these counters are often set at 'common for human' values. Although many are likely to be set to small values, we knew that some counters would be set to around 1000 (e.g. tracing in RPython VMs such as PyPy occurs at 1039 iterations of a loop) and that we might observe VM effects at in-process iterations around this point. With that in mind, it was then necessary to run benchmarks significantly past 1000 in-process iterations so that we would be able to observe any long-term effects.

To understand whether we could have achieved similar quality results with fewer in-process iterations, we truncated all of $\text{Linux}_{1240v5}$'s process executions (taking in-process iterations from $0 \ldots n$ where $n \in \{10, 20, \ldots, 1980, 1990\}$), repeated the changepoint analysis, and compared the resulting statistics (the classifications, times to reach a steady state, and steady state performances) might cover the 'true' mean while the original ⟨VM, benchmark⟩ pair does not, and sometimes neither will cover the 'true' mean. As we are unable to tell when this happens, any such comparison will still be counted against the sampled ⟨VM, benchmark⟩ pair.

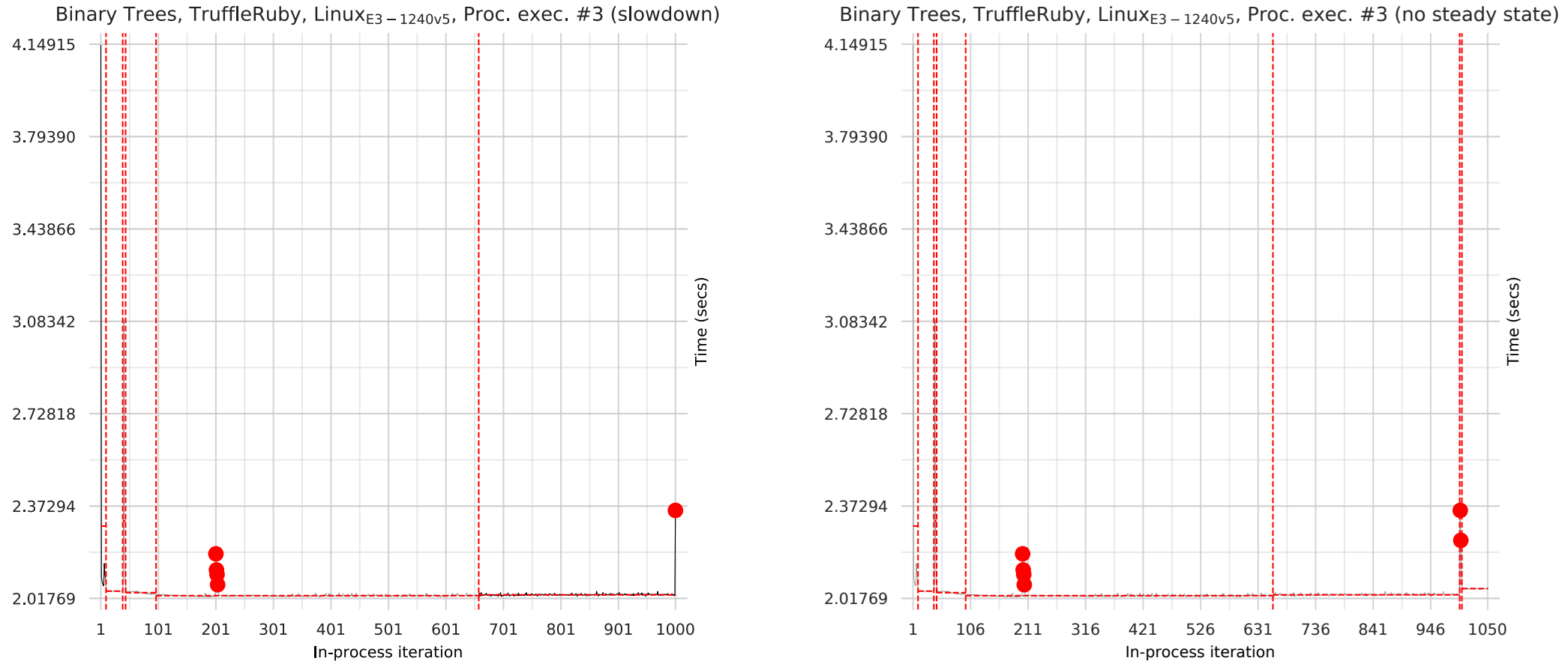

Fig. 10. Comparing the changepoints and classifications of a single JRuby+Truffle process execution originally classified as warmup when truncated at $n = 1000$ (LHS) and $n = 1050$ (RHS). Many JRuby+Truffle process executions noticeably slowdown around in-process iteration 1050 before speeding back up later, often leading to a changepoint segment being identified. Since this period is entirely missed at $n = 1000$, the resulting classification is the same as at $n = 2000$. At $n = 1050$, enough of this period is present to result in a new changepoint segment, which then causes a classification of no steady state.

to the original ($n = 2000$). We shrank the sliding window and steady state length in proportion to the number of iterations (i.e. at $n = 1000$ the sliding window is 100 and steady state length 250).

Intuitively, one would expect the similarity to decrease as $n$ reduces, allowing one to pick a value of $n$ that has a good trade-off between accuracy and running time. As Figure 9 shows, there can be significant kinks, and lengthy plateaus, in the data that make choosing such a value tricky. For example, probably the most notable kink is in the classifications similarity between $1000 \le n \le 1300$ (which then has a noticeable effect on the 'overall' similarity). This is almost entirely caused by JRuby+Truffle, many of whose process executions noticeably slow down around in-process iteration 1050 before speeding back up later, often leading to a changepoint segment being identified. At $n = 2000$, changepoint analysis considers this slower segment to be equivalent to the final segment (generally leading to a classification of slowdown or warmup). At $n = 1000$ the segment is not detected at all, generally leading to the same classification as at $n = 2000$. At $n = 1050$, however, a segment is detected. Since it is the process execution's final segment, and since it typically has a smaller mean and variance than the 'real' last segment, it is often considered non-equivalent to the preceding segment, which then often leads to a classification of no steady state (see Figure 10 for an example).

It is thus difficult to draw definitive conclusions from our results: for example, the 'overall' similarity is better at $n = 700$ (85%) than it is at $n = 900$ (78%). It is clear that the quality of results from reducing the number of in-process executions is heavily dependent on the VM(s) involved: if one is confident that the VM(s) do not cause unpredictable performance changes, then it may be possible to run many fewer in-process iterations; if one is unsure then running more is the only safe course of action. We suggest that in many cases a reasonable compromise might be to use smaller numbers (e.g. 500) of in-process iterations most of the time, while occasionally using larger numbers (e.g. 1500) to see if longer-term stability has been affected.

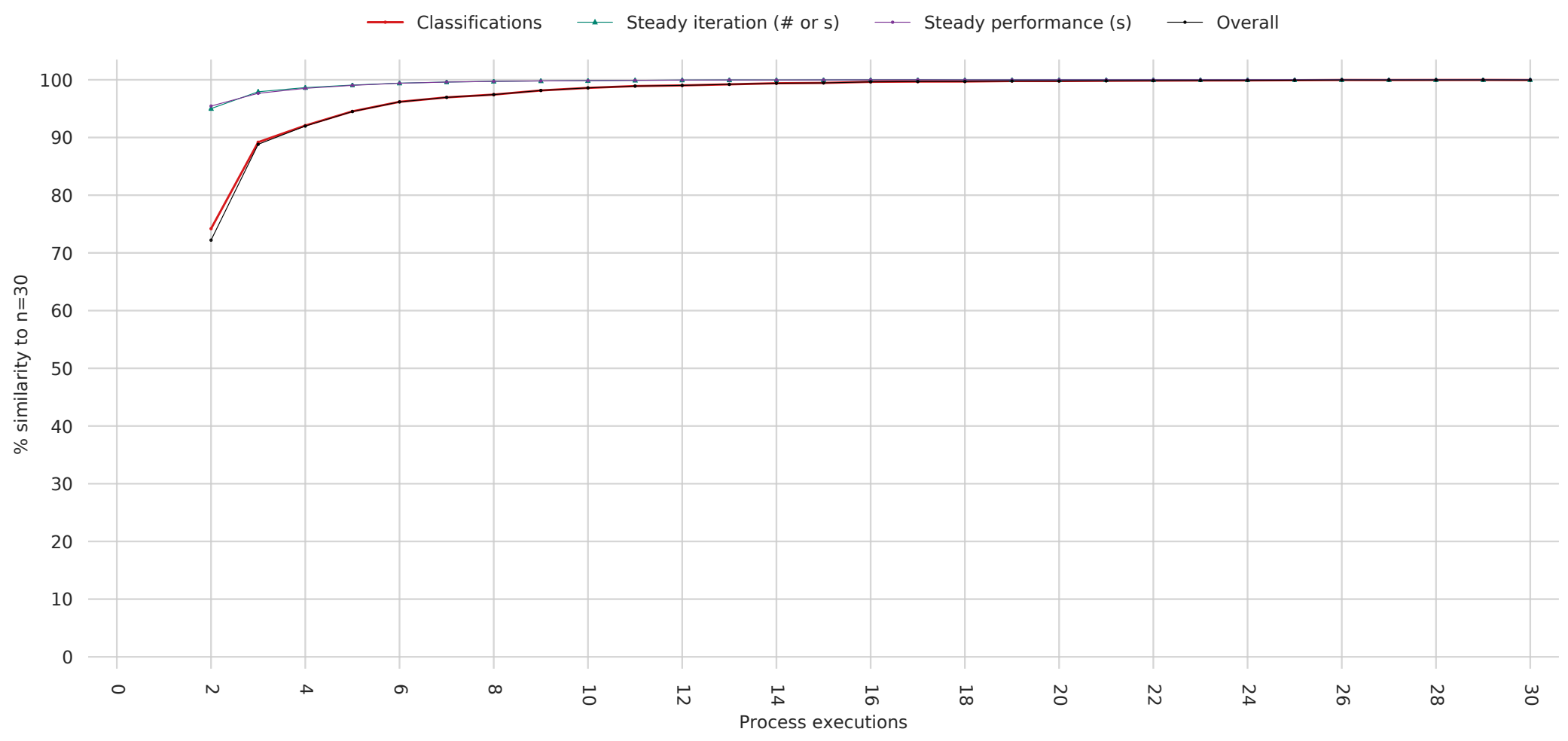

Fig. 11. How similar $\text{Linux}_{1240\text{v}5}$'s results (seen in Figure 2) would have been if we had run fewer than 30 process executions. For each ⟨VM, benchmark⟩ pair and each $n \in \{2, 3, \ldots, 29, 30\}$ we created 1000 bootstrapped samples (using sampling with replacement) of size $n$ (i.e. each set contains $n$ process executions) and compared their classifications, time to reach a steady state, and steady state performance to the original ($n = 30$). The 'overall' similarity of each truncated dataset is the number of ⟨VM, benchmark⟩ pairs for which all three characteristics are statistically equivalent to the full experiment. In general, lower values of $n$ lead to fewer ⟨VM, benchmark⟩ pairs being similar to $n = 30$.

*9.1.2 Process executions.* We had little intuition as to how many process executions we should run, so our eventual choice of 30 was somewhat arbitrary. To understand if we could have used fewer process executions, we used a slightly different method to the analysis of in-process iterations. For each ⟨VM, benchmark⟩ pair we created 1000 bootstrapped sample sets (using sampling with replacement) of size $n \in \{2, 3, \ldots, 29, 30\}$, from the original $\text{Linux}_{1240\text{v}5}$ experiment results and compared the resulting statistics (the classifications, times to reach a steady state, and steady state performances) to the original. Figure 11 shows the results of this analysis.

Our analysis shows the importance of running multiple process executions: at $n = 2$, for example, the quality of results is relatively poor, particularly for classifications; but at $n = 6$ one can clearly observe the beginning of the point of diminishing returns. We would have had very similar results to our final experiment with just 10 process executions and all but identical results with 15 process executions. This suggests that other benchmarking may be able to get good results from 10 process executions though, as with in-process iterations, occasionally running larger numbers of process executions will help identify infrequent performance issues.

## 10 CONCLUSIONS

Warmup has previously been an informally defined term [Seaton 2015]. We captured a restricted version of this definition in Hypothesis H1, which a carefully designed experiment then invalidated. The analysis techniques we used are open-source and can be used by VM developers and users to investigate the warmup behaviour of any VM and benchmark.

Although we are fairly experienced in designing and implementing experiments, the experiment in this paper took far more time than we expected — around 3 person years. In part this is because there is limited precedent for such detailed experiments. Investigating possible confounding variables, understanding how to control them, and implementing the necessary checks, all took time. In many cases, we had to implement small programs or systems to understand a variable's effect (e.g. that Linux allows a process to allocate memory beyond that specified in the soft and hard `ulimit`). However, we are realistic that few people will have the time or energy to institute all the controls that we implemented. An open question is which of the controls are the most significant in terms of producing a reliable experiment. The large number of partly inter-locking combinations means that we estimate that untangling this will require significant additional running time (at least many months).

## ACKNOWLEDGMENTS

We are particularly grateful to Vincent Knight who helped put together this paper's team. Chris Seaton helped integrate JRuby+Truffle into our experiment. Lukas Diekmann rebooted more servers than any person should have to. We also thank (in alphabetical order) Dave Dice, Kenny Gross, Tim Harris, Tomas Kalibera, Ben Titzer, Mircea Trofin, Dave Ungar, and Mario Wolczko for comments and suggestions; any errors and infelicities are our own. This research was funded by the EPSRC Cooler (EP/K01790X/1) grant and Lecture (EP/L02344X/1) fellowship, and a gift from Oracle Labs.

**Transparency:** Three of this paper's authors have contributed to the PyPy project, and one is an OpenBSD developer. The King's group has received several funding gifts from Oracle Labs.

In this appendix, we first show that our statistical method can be applied to well known benchmarking suites (Appendix A) before showing additional results from the main part of our experiment (Appendix B). We then present a curated selection of interesting run-sequence plots (Appendix C). The complete series of plots is available in a separate document.

## A APPLYING THE STATISTICAL METHOD TO EXISTING BENCHMARK SUITES

The statistical method presented in Section 4 is not limited to data produced from Krun. To demonstrate this, we have applied it to two standard benchmark suites: the DaCapo suite for Java [Blackburn et al. 2006] and the Octane suite for JavaScript [Google 2012]. Octane was run on all three of our benchmarking machines, whereas (due to time constraints), Dacapo was run only on $\text{Linux}_{4790}$. For both suites we used 30 process executions and 2000 in-process iterations.

We ran DaCapo (with its default benchmark size) on Graal and HotSpot. As it already has support for altering the number of in-process iterations, we used it without modification. However, we were unable to run 3 of its 14 benchmarks: `batik` crashes with a `InvocationTargetException`; `eclipse`, `tomcat` fail their own internal validation checks. We experienced semi-regular hangs with Graal which led us to disable `luindex`, `tradebeans`, and `tradesoap` on Graal only.

We ran Octane on the same version of V8 used in the main experiment, and (on Linux only, due to benchmark failures on OpenBSD) on SpiderMonkey (#1196bf3032e1, a JIT compiling VM for JavaScript). We replaced its complex runner (which reported timings with a non-monotonic microsecond timer) with a simpler alternative (using a monotonic millisecond timer). We also had to decide on an acceptable notion of 'iteration'. Many of Octane's benchmarks consist of a relatively quick 'inner benchmark'; an 'outer benchmark' specifies how many times the inner benchmark should be run in order to make an adequately long in-process iteration. We recorded 2000 in-process iterations of the outer benchmark; our runner fully resets the benchmark and the random number generator between each iteration. The `box2d`, `gameboy`, `mandreel` benchmarks do not properly reset their state between runs, leading to run-time errors we have not been able to fix; `typescript`'s reset function, in contrast, frees constant data needed by all iterations, which we were able to easily fix. When run for 2000 iterations, `CodeLoadClosure`, `pdfjs`, and `zlib` all fail due to memory leaks. We were able to easily fix `pdfjs` by emptying a global list after each iteration, but not the others. We therefore include 12 of Octane's benchmarks (including lightly modified versions of `pdfjs` and `typescript`). Because we run fewer benchmarks, our modified runner is unable to fully replicate the running order of Octane's original runner. Since Octane runs all benchmarks in a single process execution, this could affect the performance of later benchmarks in the suite.

Table 4 shows the full DaCapo results. Because we had to run a subset of benchmarks on Graal, a comparison with HotSpot is difficult though of the benchmarks both could run, there is a reasonable degree of similarity. However, even on this most carefully designed of benchmark suites, only 42% of ⟨VM, benchmark⟩ pairs have 'good' warmup.

Tables 5–7 show the full Octane results. These show a greater spread of classifications than the DaCapo results, with 33% of ⟨VM, benchmark⟩ pairs over the three machines having 'good' warmup.

As these results show, our automated statistical method produces satisfying results even on existing benchmark suites that have not been subject to the Krun treatment. Both DaCapo and (mostly) Octane use much larger benchmarks than our main experiment. We have no realistic way of understanding to what extent this makes 'good' warmup more or less likely. For example, it is likely that there is CFG non-determinism in many of these benchmarks; however, their larger code-bases may give VMs the ability to 'spread out' VM costs, making smaller blips less noticeable.

Table 4. DaCapo results for $\text{Linux}_{4790}$.

| | Graal Class. | Graal Steady iter (#) | Graal Steady iter (s) | Graal Steady perf (s) | Hotspot Class. | Hotspot Steady iter (#) | Hotspot Steady iter (s) | Hotspot Steady perf (s) |
|---|---|---|---|---|---|---|---|---|
| avrora | ⤨ (27⅂, 3ʃ) | 927.0 (3.0, 1103.2) | 1746.32 (5.010, 2092.957) | 1.87639 ±0.026359 | ⤨ (22⅂, 8ʃ) | 728.5 (2.0, 1057.3) | 1353.30 (2.210, 1971.490) | 1.85568 ±0.044805 |
| fop | ⤨ (27ʃ, 3⅂) | 1277.5 (136.6, 1454.8) | 178.72 (21.938, 201.865) | 0.13727 ±0.003089 | ⤨ (22ʃ, 8~) | | | |
| h2 | — | | | 3.23273 ±0.079024 | — | | | 3.00628 ±0.103632 |
| jython | ⤨ (14~, 10⅂, 6ʃ) | | | | ⤨ (14⅂, 9~, 7ʃ) | | | |
| luindex | | | | | ⤨ (21⅂, 5~, 4ʃ) | | | |
| lusearch | ⅂ | 100.0 (89.0, 101.5) | 67.78 (60.256, 69.551) | 0.65195 ±0.006567 | ⅂ | 89.0 (87.0, 106.5) | 46.05 (44.203, 54.515) | 0.50350 ±0.024075 |
| pmd | ⤨ (29⅂, 1~) | | | | ⅂ | 1344.0 (721.4, 1375.3) | 879.30 (483.471, 910.551) | 0.64894 ±0.016645 |
| sunflow | ⤨ (24⅂, 4~, 2ʃ) | | | | — | | | 1.70203 ±0.022747 |
| tradebeans | | | | | ⅂ | 3.0 (2.0, 807.2) | 5.40 (3.145, 1738.590) | 2.14287 ±0.063497 |
| tradesoap | | | | | — | | | 2.70721 ±0.039574 |
| xalan | ⤨ (20⅂, 7—, 3ʃ) | 78.0 (1.0, 569.9) | 40.72 (0.000, 270.938) | 0.46849 ±0.011542 | ⤨ (28⅂, 1~, 1ʃ) | | | |

Table 5. Octane results for Linux$_{4790}$.

| | | SpiderMonkey Class. | Steady iter (#) | Steady iter (s) | Steady perf (s) | | V8 Class. | Steady iter (#) | Steady iter (s) | Steady perf (s) |
|---|---|---|---|---|---|---|---|---|---|---|
| Boyer | SpiderMonkey | ≍ (28–, 1∿, 1ʃ) | | | | V8 | ≍ (13∿, 11ɿ, 6ʃ) | | | |
| Decrypt | | ≍ (11–, 11∿, 8ɿ) | | | | | ≍ (22ɿ, 4–, 3∿, 1ʃ) | | | |
| DeltaBlue | | ≍ (15–, 6ɿ, 5∿, 4ʃ) | | | | | = (29ɿ, 1–) | 3.0 (3.0, 5.1) | 0.94 (0.920, 1.917) | 0.45343 ±0.012796 |
| Earley | | ≍ (14–, 7ʃ, 6∿, 3ɿ) | | | | | ≍ (18∿, 8ɿ, 3ʃ, 1–) | | | |
| Encrypt | | ≍ (21–, 8ɿ, 1ʃ) | 1.0 (1.0, 4.0) | 0.00 (0.000, 2.388) | 0.77672 ±0.018463 | | ≍ (19ɿ, 9–, 2ʃ) | 2.0 (1.0, 39.3) | 0.81 (0.000, 30.462) | 0.79224 ±0.011236 |
| NavierStokes | | ɿ | 2.0 (2.0, 23.5) | 0.76 (0.753, 15.658) | 0.69113 ±0.001010 | | = (25ɿ, 5–) | 2.0 (1.0, 4.6) | 0.91 (0.000, 3.282) | 0.88885 ±0.004485 |
| PdfJS | | ɿ | 2.0 (2.0, 333.1) | 1.44 (1.423, 394.672) | 1.19056 ±0.026962 | | ≍ (16ʃ, 12ɿ, 2∿) | | | |
| RayTrace | | ≍ (16ɿ, 9∿, 3–, 2ʃ) | | | | | ≍ (29ɿ, 1∿) | | | |
| RegExp | | = (25–, 5ɿ) | 1.0 (1.0, 10.0) | 0.00 (0.000, 9.792) | 1.05061 ±0.022040 | | ≍ (22–, 7ɿ, 1ʃ) | 1.0 (1.0, 243.7) | 0.00 (0.000, 208.640) | 0.82839 ±0.025454 |
| Richards | | ≍ (25ɿ, 5ʃ) | 2.0 (2.0, 1111.3) | 0.83 (0.830, 892.658) | 0.80609 ±0.002301 | | ≍ (27ɿ, 3∿) | | | |
| Splay | | = (27–, 3ɿ) | 1.0 (1.0, 202.7) | 0.00 (0.000, 145.962) | 0.67068 ±0.018679 | | — | | | 0.52488 ±0.006651 |
| Typescript | | — | | | 2.26833 ±0.048122 | | ɿ | 2.0 (2.0, 2.0) | 2.66 (2.580, 2.689) | 1.88786 ±0.028842 |

Table 6. Octane results for $OpenBSD_{4790}$.

| | | Class. | Steady iter (#) | Steady iter (s) | Steady perf (s) |
|---|---|---|---|---|---|
| V8 | Boyer | ≍ (16⅂, 14∿) | | | |
| | Decrypt | ≍ (18∿, 10⅂, 2ʃ) | | | |
| | DeltaBlue | ≍ (29⅂, 1∿) | | | |
| | Earley | ≍ (25∿, 4ʃ, 1⅂) | | | |
| | Encrypt | ≍ (21⅂, 8—, 1ʃ) | 5.0 (1.0, 90.0) | 3.17 (0.000, 70.168) | 0.78813 ±0.005147 |
| | NavierStokes | = (28⅂, 2—) | 32.5 (1.5, 101.7) | 28.10 (0.407, 89.626) | 0.88864 ±0.004125 |
| | PdfJS | ≍ (21⅂, 6ʃ, 3∿) | | | |
| | RayTrace | ≍ (29⅂, 1∿) | | | |
| | RegExp | ≍ (23—, 4⅂, 3ʃ) | 1.0 (1.0, 206.8) | 0.00 (0.000, 214.215) | 1.03212 ±0.073943 |
| | Richards | ∿ | | | |
| | Splay | — | | | 0.56815 ±0.009868 |
| | Typescript | ⅂ | 2.0 (2.0, 3.0) | 3.14 (3.056, 5.114) | 1.99426 ±0.042858 |

Table 7. Octane results for $\text{Linux}_{1240v5}$.

| | SpiderMonkey | | | | V8 | | | |
|---|---|---|---|---|---|---|---|---|
| | Class. | Steady iter (#) | Steady iter (s) | Steady perf (s) | Class. | Steady iter (#) | Steady iter (s) | Steady perf (s) |
| Boyer | ≭ (22–, 4∿, 3ʃ, 1ʅ) | | | | ≭ (15ʅ, 9ʃ, 6∿) | | | |
| Decrypt | ≭ (14ʅ, 9ʃ, 5–, 2∿) | | | | = (25ʅ, 5–) | 4.0 (1.0, 10.0) | 2.40 (0.000, 7.128) | 0.78917 ±0.012041 |
| DeltaBlue | ≭ (26–, 2ʃ, 1∿, 1ʅ) | | | | ≭ (17ʅ, 8∿, 4ʃ, 1–) | | | |
| Earley | ≭ (27–, 3ʃ) | 1.0 (1.0, 670.7) | 0.00 (0.000, 824.615) | 1.23534 ±0.009218 | ≭ (14∿, 11ʅ, 5ʃ) | | | |
| Encrypt | ʅ | 2.0 (2.0, 871.9) | 0.77 (0.765, 612.432) | 0.70199 ±0.002665 | = (29ʅ, 1–) | 2.0 (2.0, 16.5) | 0.81 (0.805, 12.230) | 0.78636 ±0.006621 |
| NavierStokes | ʅ | 2.0 (2.0, 22.2) | 0.85 (0.847, 16.750) | 0.78570 ±0.001705 | = (28ʅ, 2–) | 2.0 (1.5, 3.6) | 0.98 (0.437, 2.588) | 0.96251 ±0.003949 |
| PdfJS | ≭ (29ʅ, 1ʃ) | 2.0 (2.0, 231.6) | 1.39 (1.375, 261.885) | 1.14052 ±0.036265 | ≭ (18ʅ, 11ʃ, 1∿) | | | |
| RayTrace | ≭ (17ʅ, 8∿, 3ʃ, 2–) | | | | ≭ (29ʅ, 1∿) | | | |
| RegExp | ≭ (12∿, 11ʃ, 4ʅ, 3–) | | | | ≭ (22–, 4ʅ, 4ʃ) | 1.0 (1.0, 15.1) | 0.00 (0.000, 11.288) | 0.79499 ±0.022808 |
| Richards | ≭ (27ʅ, 3∿) | | | | ≭ (25ʅ, 5∿) | | | |
| Splay | = (22–, 8ʅ) | 1.0 (1.0, 336.3) | 0.00 (0.000, 216.954) | 0.59816 ±0.013103 | – | | | 0.46021 ±0.008743 |
| Typescript | = (29–, 1ʅ) | 1.0 (1.0, 1.0) | 0.00 (0.000, 0.000) | 1.99680 ±0.030530 | ≭ (29ʅ, 1ʃ) | 2.0 (2.0, 3.0) | 2.38 (2.316, 4.019) | 1.64553 ±0.018712 |

## B FURTHER RESULTS

The main experiment's results for $\text{Linux}_{4790}$ and $\text{OpenBSD}_{4790}$ can be seen in Tables 8 and 9.

Table 8. Benchmark results for $\text{Linux}_{4790}$.

| | | Class. | Steady iter (#) | Steady iter (s) | Steady perf (s) | | Class. | Steady iter (#) | Steady iter (s) | Steady perf (s) |
|---|---|---|---|---|---|---|---|---|---|---|
| C | *binary trees* | ∿ | | | | *n-body* | ≍ (15–, 7∿, 4ʃ, 4ʅ) | | | |
| Graal | | ʃ | 133.0 (121.2, 343.8) | 25.84 (23.577, 65.918) | 0.19044 ±0.000675 | | ʅ | 6.0 (6.0, 6.0) | 1.03 (1.027, 1.042) | 0.16717 ±0.000032 |
| HHVM | | ≍ (18ʅ, 8ʃ, 4∿) | | | | | ʅ | 135.0 (132.0, 169.5) | 1170.06 (1148.762, 1277.609) | 3.14291 ±0.031743 |
| HotSpot | | ʅ | 5.0 (2.0, 10.0) | 0.84 (0.247, 1.816) | 0.19004 ±0.001142 | | ʅ | 2.0 (2.0, 2.0) | 0.18 (0.177, 0.178) | 0.17382 ±0.000053 |
| JRuby+Truffle | | ≍ (19ʅ, 7ʃ, 4∿) | | | | | ʅ | 69.0 (68.0, 69.0) | 21.18 (20.840, 21.473) | 0.25414 ±0.001788 |
| LuaJIT | | ≍ (23–, 5ʅ, 1∿, 1ʃ) | | | | | – | | | 0.29588 ±0.003017 |
| PyPy | | ≍ (20ʃ, 10∿) | | | | | ≍ (29–, 1ʃ) | 1.0 (1.0, 1.0) | 0.00 (0.000, 0.000) | 1.83351 ±0.097657 |
| V8 | | ≍ (13∿, 9ʃ, 7ʅ, 1–) | | | | | – | | | 0.28104 ±0.000413 |
| C | *fannkuch redux* | ≍ (27–, 2ʅ, 1∿) | | | | *Richards* | ≍ (24–, 2ʃ, 2ʅ, 2∿) | | | |
| Graal | | ≍ (27ʅ, 3∿) | | | | | ≍ (29ʅ, 1ʃ) | 2.0 (2.0, 972.9) | 1.07 (1.049, 272.439) | 0.28029 ±0.005656 |
| HHVM | | ≍ (29ʅ, 1ʃ) | 10.0 (10.0, 77.2) | 52.37 (52.340, 152.031) | 1.41340 ±0.010172 | | | | | |
| HotSpot | | ≍ (18ʅ, 11∿, 1ʃ) | | | | | ≍ (28ʃ, 2∿) | | | |
| JRuby+Truffle | | ʃ | 1016.5 (999.0, 1023.5) | 1060.10 (1040.530, 1087.273) | 1.14610 ±0.036268 | | ʃ | 1021.0 (423.0, 1029.0) | 998.76 (422.492, 1042.619) | 0.97756 ±0.045730 |
| LuaJIT | | ≍ (29–, 1∿) | | | | | = (25–, 5ʅ) | 1.0 (1.0, 77.9) | 0.00 (0.000, 222.429) | 2.90166 ±0.039125 |
| PyPy | | = (27ʅ, 3–) | 2.0 (1.0, 49.5) | 1.60 (0.000, 76.408) | 1.56668 ±0.008621 | | ʅ | 2.0 (2.0, 60.3) | 1.20 (1.182, 61.955) | 1.04161 ±0.025098 |
| V8 | | = (27–, 3ʅ) | 1.0 (1.0, 2.0) | 0.00 (0.000, 0.309) | 0.30419 ±0.000225 | | ≍ (25ʅ, 5ʃ) | 5.5 (4.0, 793.0) | 2.54 (1.674, 440.327) | 0.55482 ±0.018615 |
| C | *fasta* | – | | | 0.07408 ±0.000159 | *spectralnorm* | ≍ (18ʅ, 8ʃ, 4–) | 2.0 (1.0, 2.0) | 0.55 (0.000, 0.642) | 0.55059 ±0.069302 |
| Graal | | ʅ | 5.0 (4.0, 7.0) | 0.98 (0.800, 1.272) | 0.14911 ±0.001320 | | ≍ (21ʃ, 9–) | 14.0 (1.0, 15.0) | 13.42 (0.000, 14.470) | 1.03519 ±0.000117 |
| HHVM | | ʅ | 20.0 (10.0, 229.9) | 18.74 (10.371, 187.641) | 0.80032 ±0.012900 | | ʅ | 37.0 (34.0, 206.2) | 137.93 (132.259, 415.345) | 1.63782 ±0.000404 |
| HotSpot | | ʅ | 7.0 (7.0, 595.0) | 0.70 (0.686, 66.664) | 0.11023 ±0.001137 | | ʅ | 9.0 (7.0, 14.0) | 3.84 (2.880, 6.228) | 0.47780 ±0.000061 |
| JRuby+Truffle | | | | | | | ≍ (28ʅ, 2ʃ) | 1007.5 (1000.8, 1012.0) | 1044.90 (1037.879, 1049.182) | 1.01299 ±0.013035 |
| LuaJIT | | ≍ (28–, 1ʅ, 1∿) | | | | | – | | | 0.46789 ±0.000070 |
| PyPy | | ∿ | | | | | ʅ | 75.0 (75.0, 75.0) | 38.46 (38.457, 38.472) | 0.51931 ±0.000090 |
| V8 | | ≍ (21ʃ, 9ʅ) | 793.0 (791.9, 793.5) | 912.64 (907.756, 929.744) | 1.15446 ±0.013060 | | ʅ | 3.0 (3.0, 3.0) | 0.94 (0.939, 0.939) | 0.46787 ±0.000076 |

Table 9. Benchmark results for $\text{OpenBSD}_{4790}$.

|  |  | Class. | Steady iter (#) | Steady iter (s) | Steady perf (s) |  | Class. | Steady iter (#) | Steady iter (s) | Steady perf (s) |
|---|---|---|---|---|---|---|---|---|---|---|
| C | *binary trees* | – | | | 5.16555<br>±0.112386 | *n-body* | ≭ (13ʅ, 8–, 6ʬ, 3ʃ) | | | |
| HotSpot | *binary trees* | ≭ (26ʅ, 4ʃ) | 2.0<br>(2.0, 256.0) | 0.26<br>(0.259, 49.727) | 0.18696<br>±0.005325 | *n-body* | ʅ | 2.0<br>(2.0, 2.0) | 0.18<br>(0.176, 0.179) | 0.17201<br>±0.000547 |
| LuaJIT | *binary trees* | ≭ (16ʅ, 6ʬ, 5–, 3ʃ) | | | | *n-body* | ≭ (19–, 9ʅ, 2ʃ) | 1.0<br>(1.0, 371.5) | 0.00<br>(0.000, 109.132) | 0.29326<br>±0.003969 |
| PyPy | *binary trees* | ≭ (21ʅ, 8–, 1ʬ) | | | | *n-body* | = (26–, 4ʅ) | 1.0<br>(1.0, 25.0) | 0.00<br>(0.000, 95.879) | 3.97051<br>±0.132457 |
| V8 | *binary trees* | ≭ (28ʅ, 1–, 1ʃ) | 236.5<br>(2.0, 243.5) | 115.39<br>(0.512, 120.764) | 0.48706<br>±0.005907 | *n-body* | = (20ʅ, 10–) | 363.0<br>(1.0, 370.5) | 101.27<br>(0.000, 104.327) | 0.27975<br>±0.002567 |
| C | *fannkuch redux* | ≭ (20–, 9ʅ, 1ʃ) | 1.0<br>(1.0, 173.4) | 0.00<br>(0.000, 71.288) | 0.41141<br>±0.004679 | *Richards* | ≭ (20–, 7ʅ, 3ʃ) | 1.0<br>(1.0, 90.2) | 0.00<br>(0.000, 68.378) | 0.76333<br>±0.017192 |
| HotSpot | *fannkuch redux* | ≭ (20ʅ, 8ʬ, 2ʃ) | | | | *Richards* | ≭ (29ʃ, 1ʬ) | | | |
| LuaJIT | *fannkuch redux* | ≭ (23–, 6ʅ, 1ʃ) | 1.0<br>(1.0, 139.6) | 0.00<br>(0.000, 78.807) | 0.56569<br>±0.000047 | *Richards* | ≭ (25ʅ, 3–, 1ʃ, 1ʬ) | | | |
| PyPy | *fannkuch redux* | ʅ | 13.0<br>(8.9, 15.0) | 32.77<br>(21.444, 38.444) | 2.69037<br>±0.019846 | *Richards* | ʅ | 21.5<br>(8.8, 58.6) | 23.76<br>(9.103, 67.611) | 1.15257<br>±0.022267 |
| V8 | *fannkuch redux* | = (22–, 8ʅ) | 1.0<br>(1.0, 80.1) | 0.00<br>(0.000, 24.093) | 0.30191<br>±0.000626 | *Richards* | ≭ (29ʅ, 1ʃ) | 6.0<br>(4.0, 211.6) | 2.79<br>(1.677, 116.997) | 0.55272<br>±0.002080 |
| C | *fasta* | – | | | 0.07417<br>±0.000154 | *spectralnorm* | ≭ (13–, 8ʅ, 7ʃ, 2ʬ) | | | |
| HotSpot | *fasta* | ≭ (19ʅ, 11ʃ) | 43.5<br>(3.0, 595.0) | 4.71<br>(0.236, 66.637) | 0.11119<br>±0.001453 | *spectralnorm* | ≭ (29ʅ, 1ʃ) | 16.0<br>(2.0, 189.7) | 7.18<br>(0.490, 90.651) | 0.47776<br>±0.000043 |
| LuaJIT | *fasta* | ≭ (20–, 9ʅ, 1ʃ) | 1.0<br>(1.0, 257.6) | 0.00<br>(0.000, 90.784) | 0.35222<br>±0.002978 | *spectralnorm* | ≭ (23–, 5ʅ, 2ʃ) | 1.0<br>(1.0, 268.8) | 0.00<br>(0.000, 125.922) | 0.46778<br>±0.000006 |
| PyPy | *fasta* | ≭ (24–, 5ʅ, 1ʃ) | 1.0<br>(1.0, 20.5) | 0.00<br>(0.000, 78.830) | 4.02068<br>±0.076419 | *spectralnorm* | ≭ (28ʅ, 2ʃ) | 43.0<br>(35.3, 188.0) | 21.85<br>(17.805, 97.475) | 0.51939<br>±0.000022 |
| V8 | *fasta* | ≭ (24ʃ, 6ʅ) | 317.0<br>(296.0, 465.8) | 366.11<br>(342.425, 539.258) | 1.16088<br>±0.019101 | *spectralnorm* | = (29ʅ, 1–) | 8.0<br>(3.0, 133.2) | 3.28<br>(0.942, 62.184) | 0.46779<br>±0.000005 |

## C CURATED PLOTS

The remainder of this appendix shows curated plots: we have selected 4 interesting plots from each classification, to give readers a sense of the range of data obtained from our experiment. A separate document contains the complete series of plots.

## C.1 Examples of Warmup Behaviour

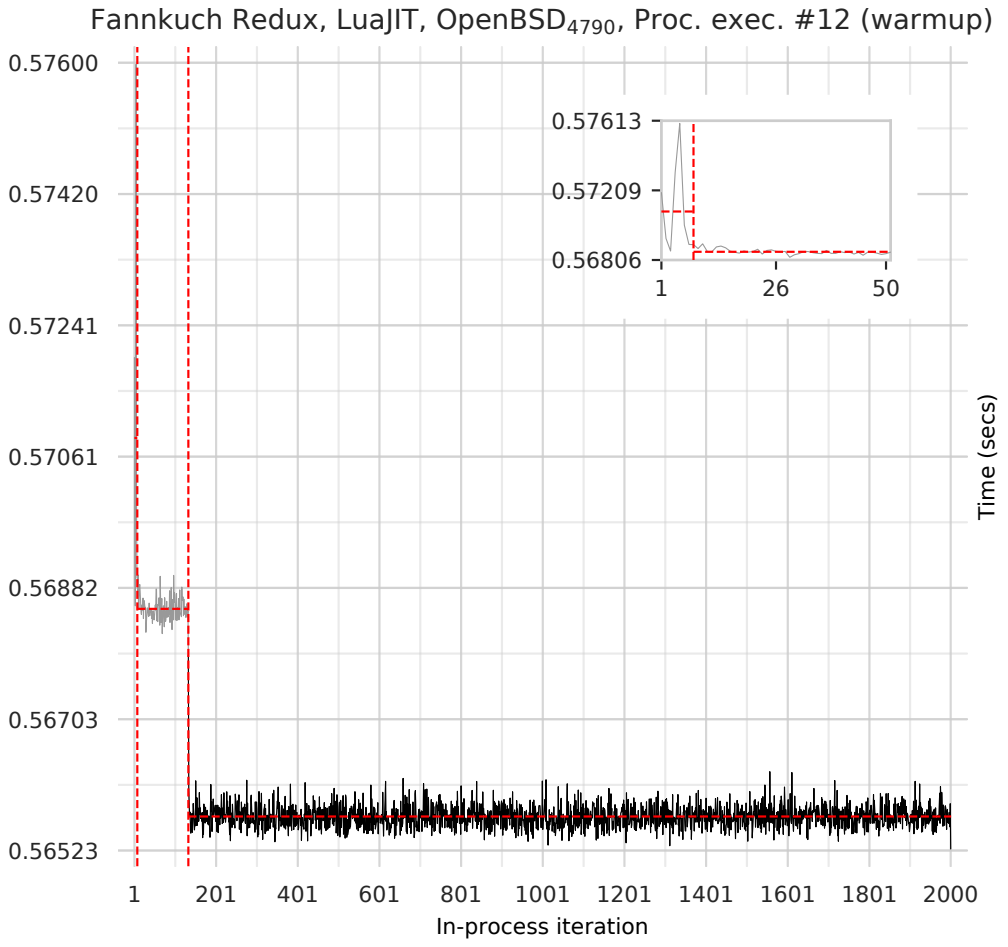

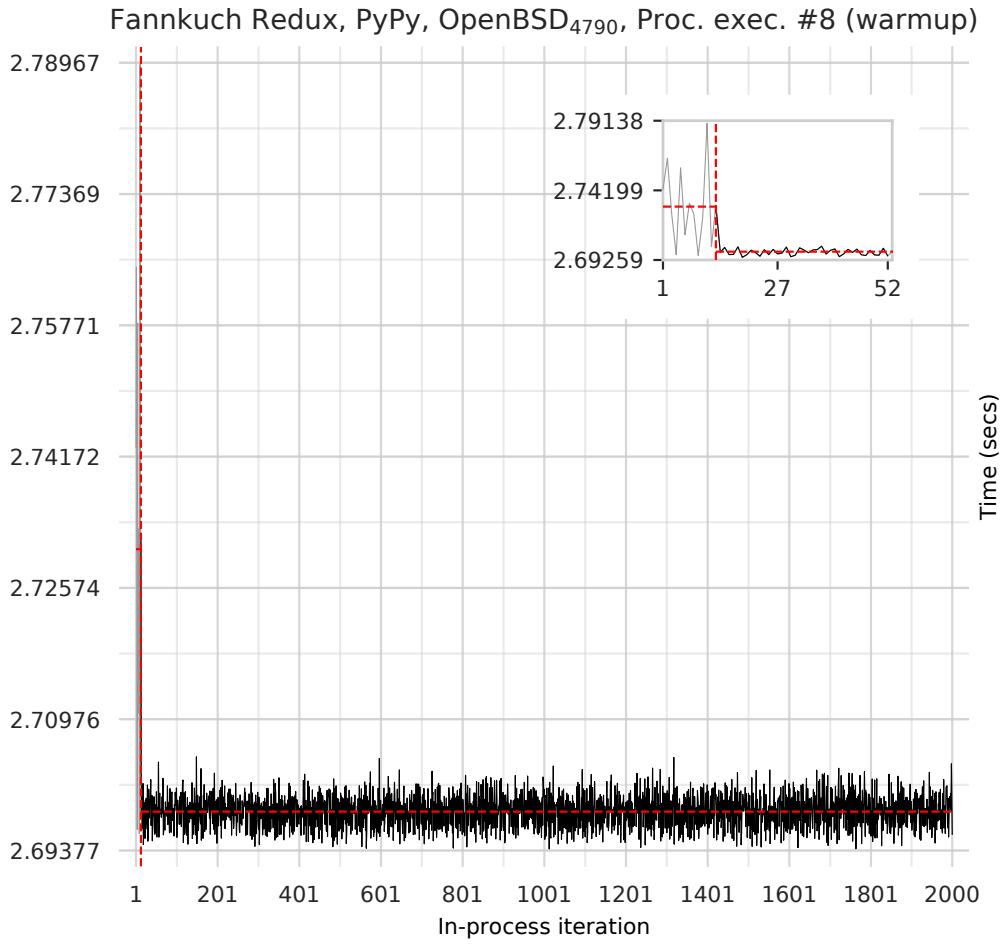

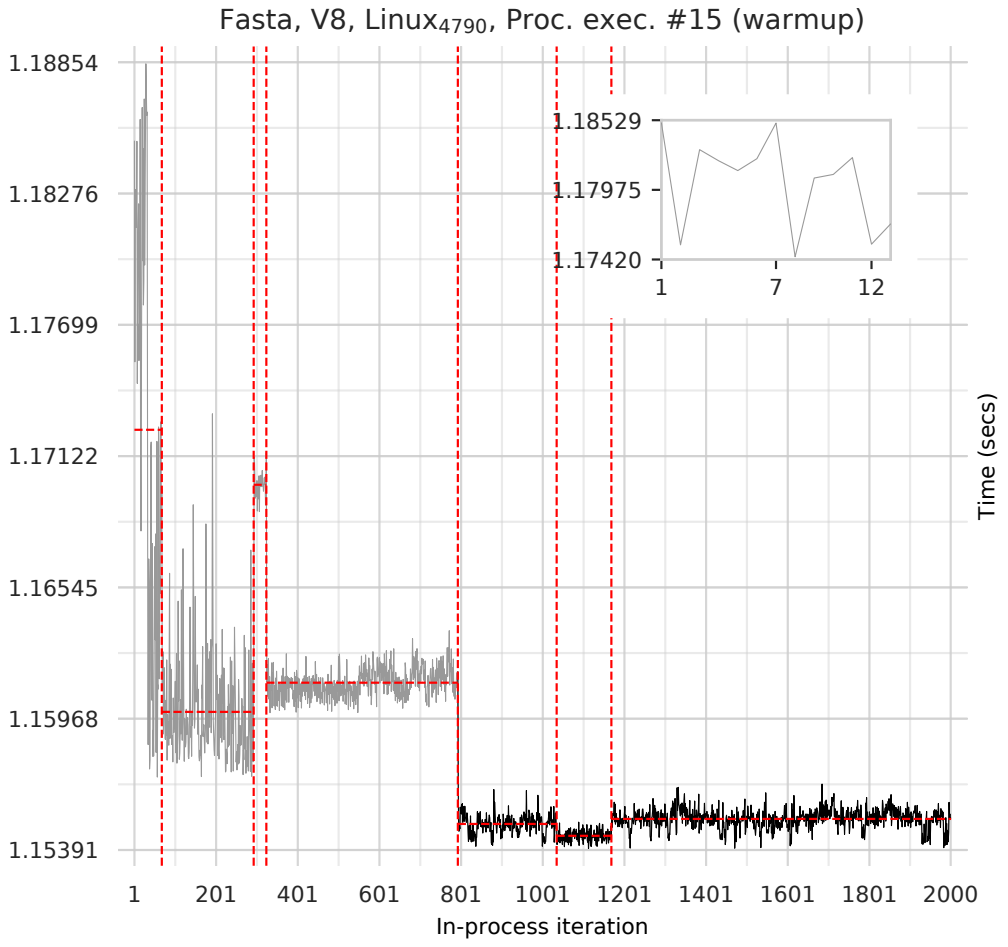

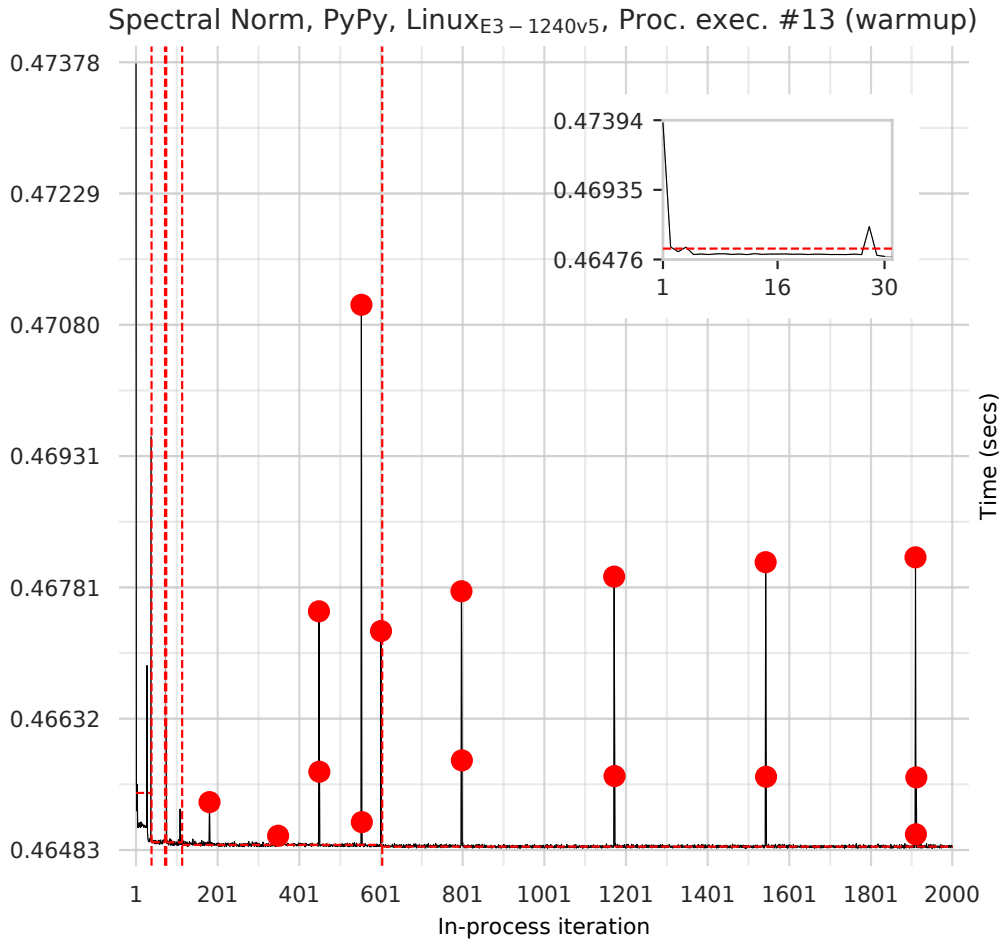

## C.2 Examples of Flat Behaviour

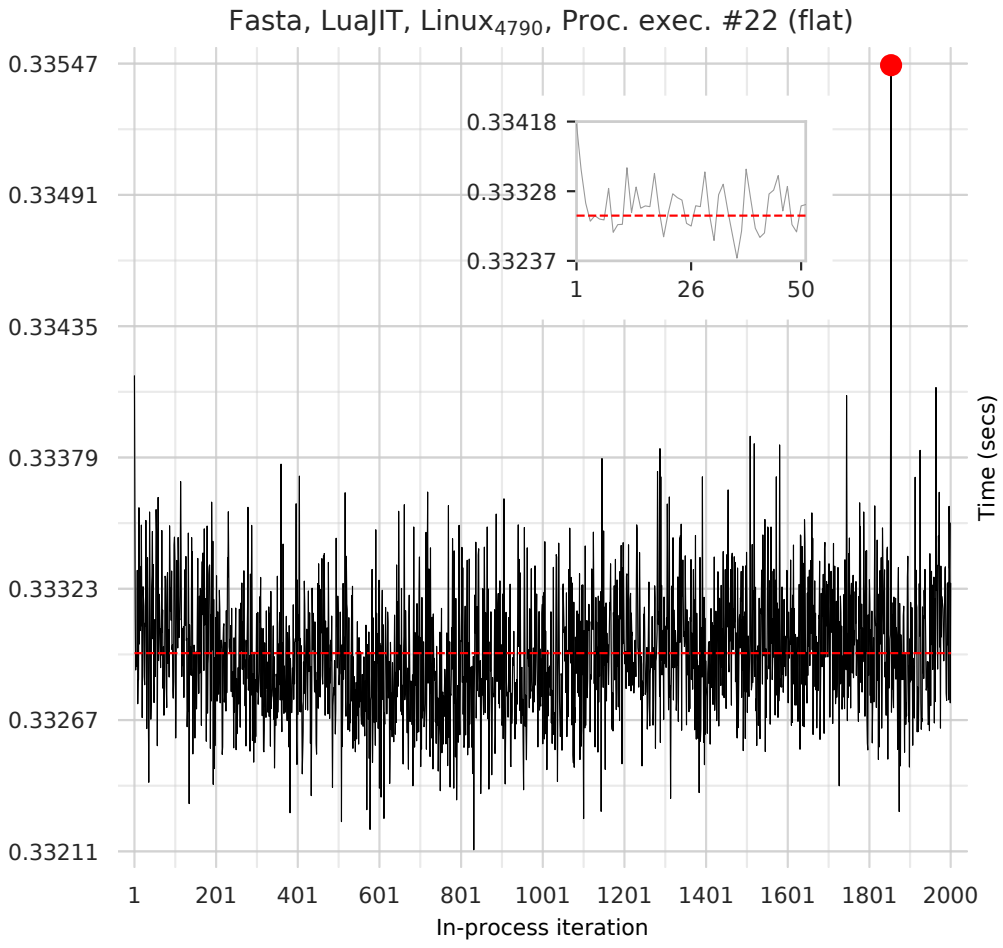

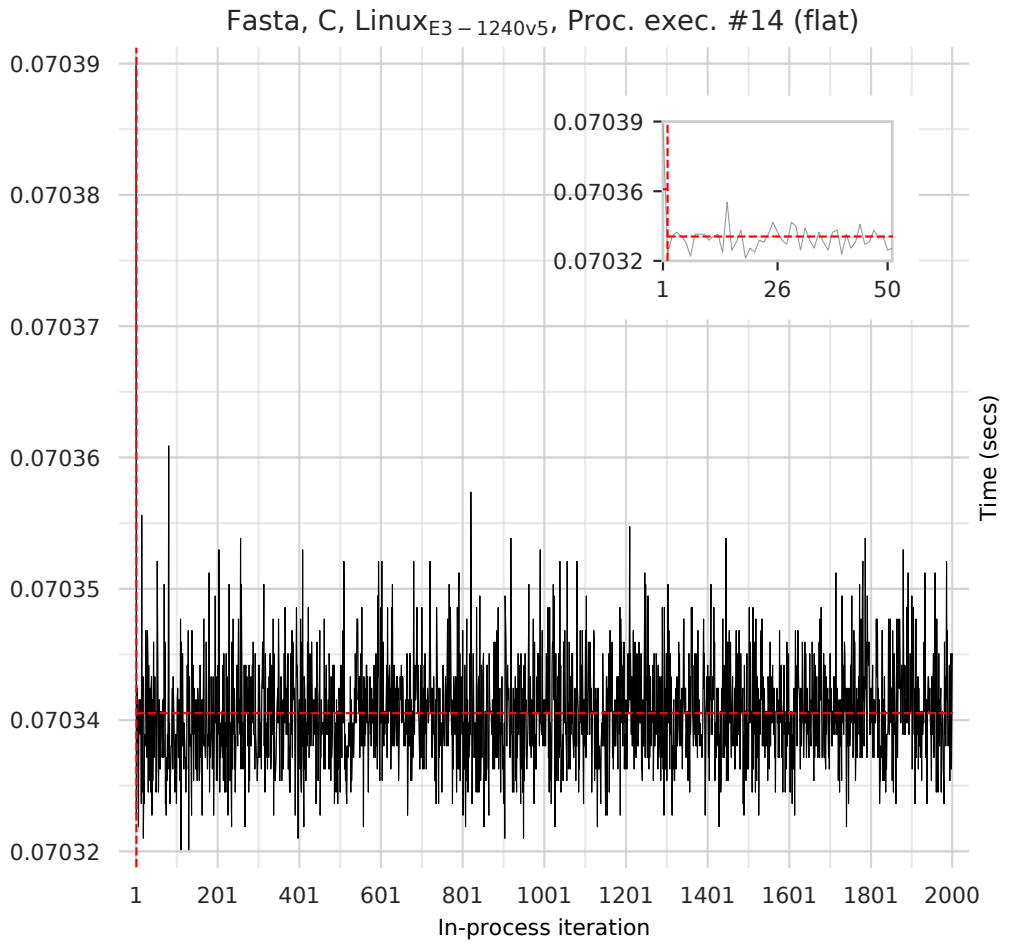

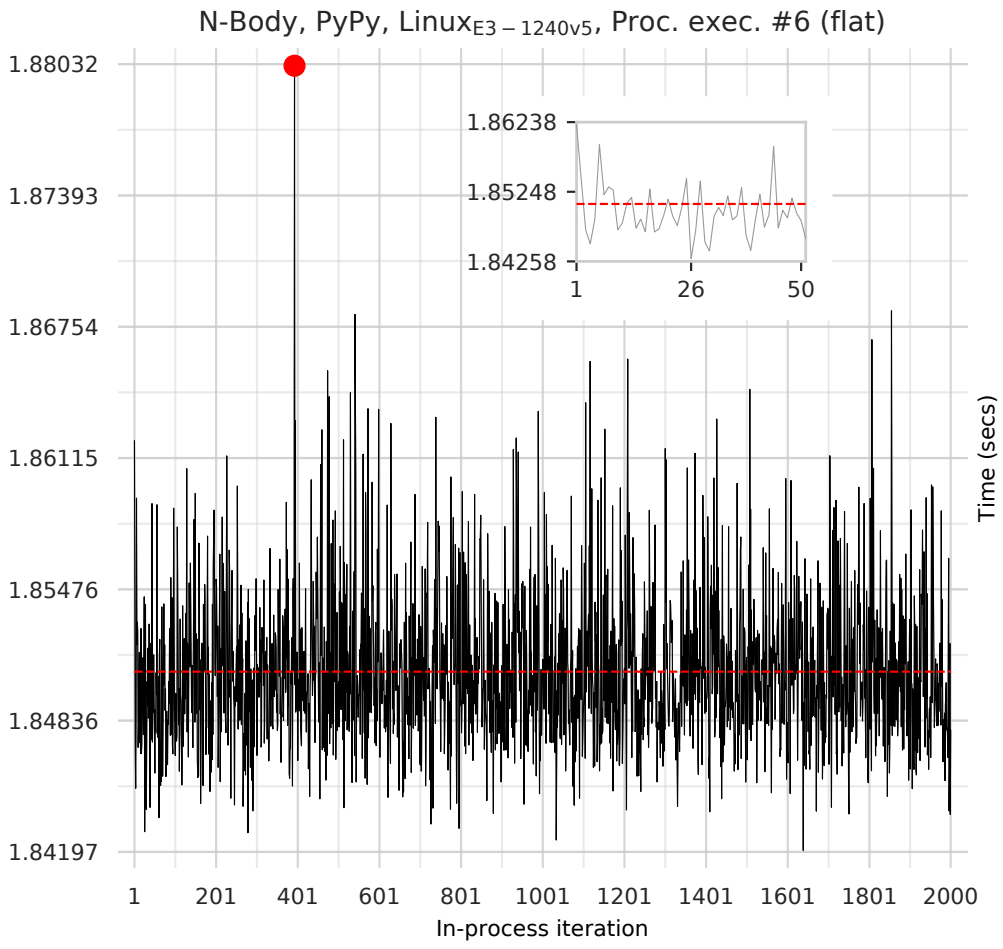

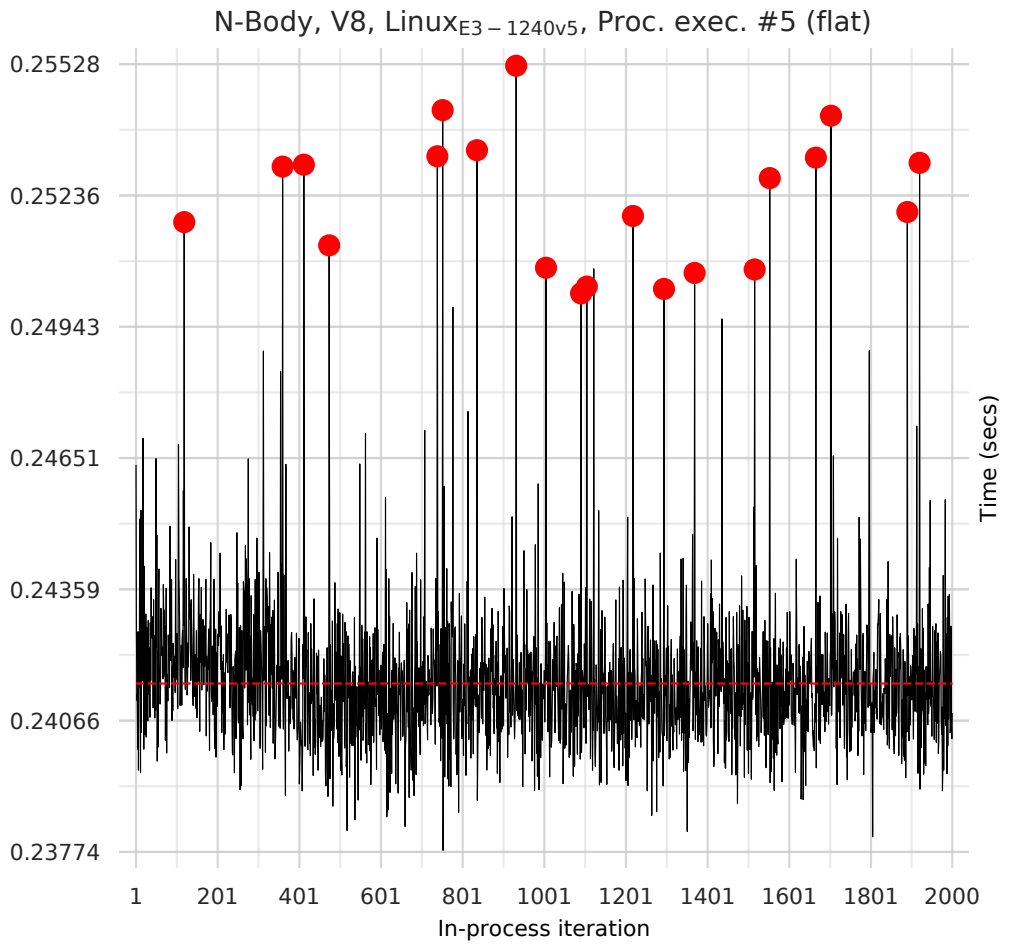

## C.3 Examples of Slowdown Behaviour

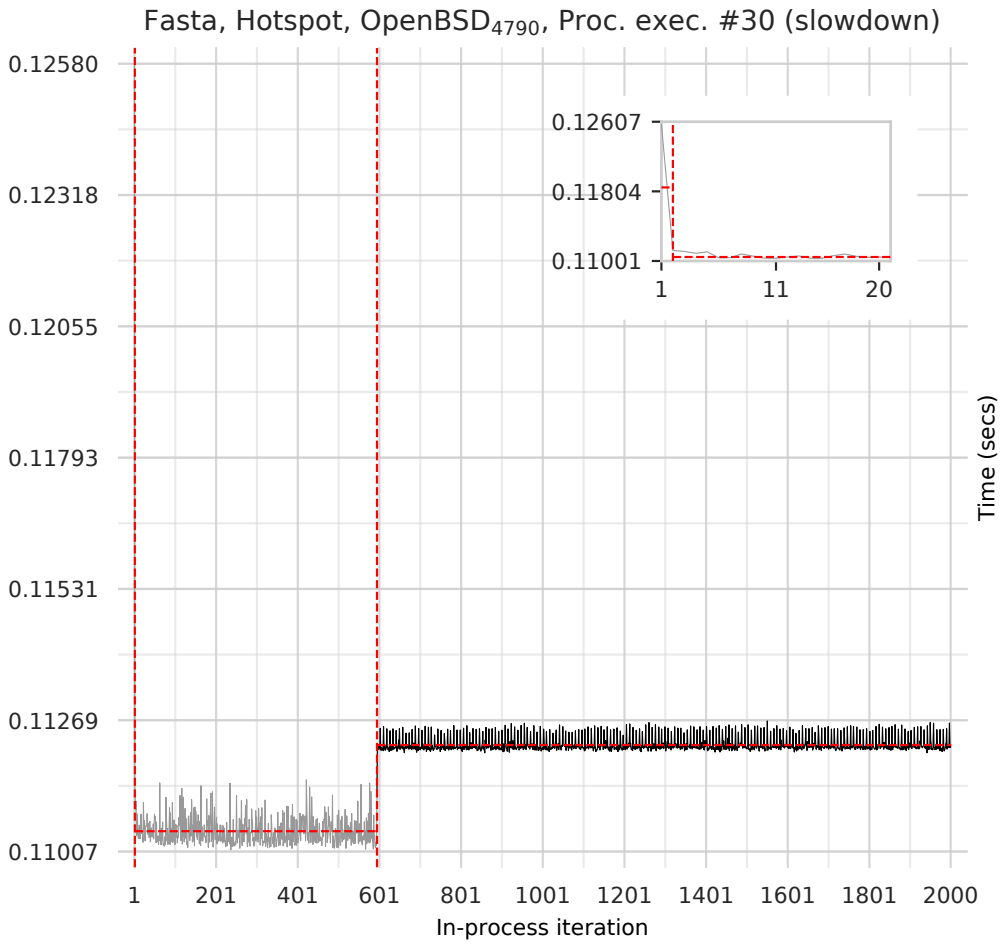

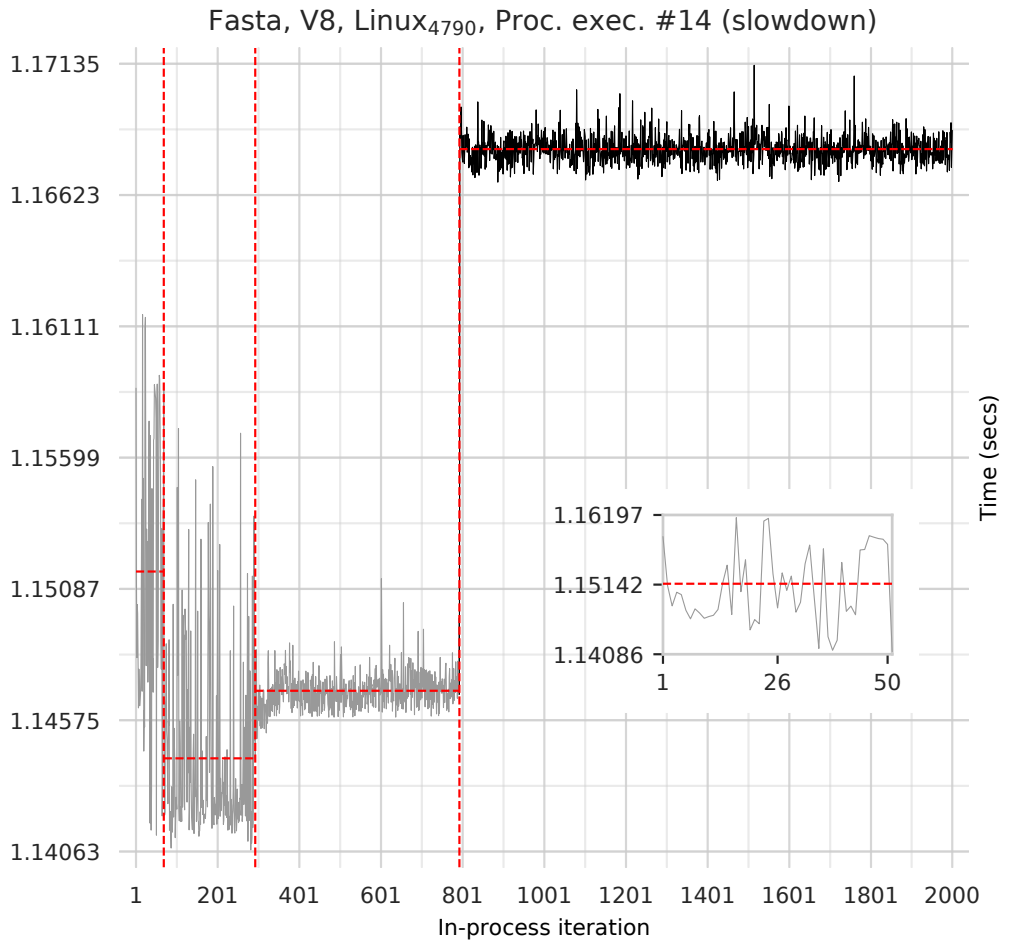

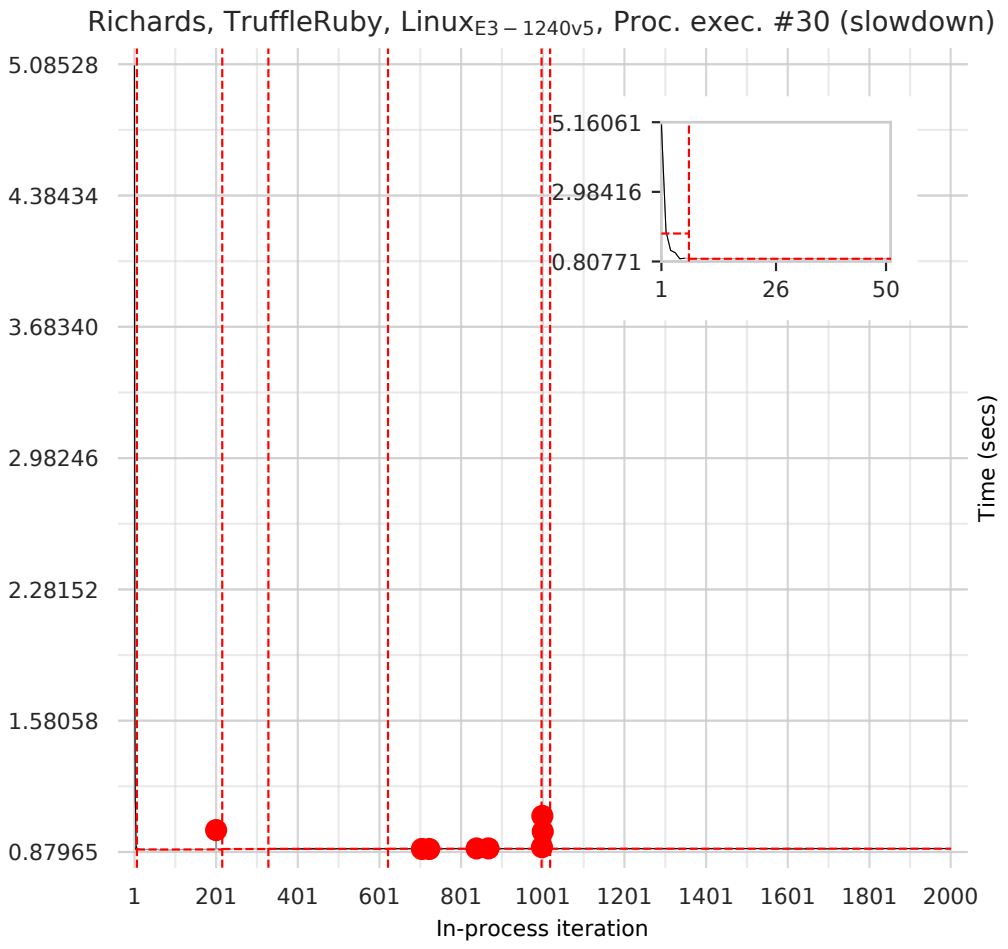

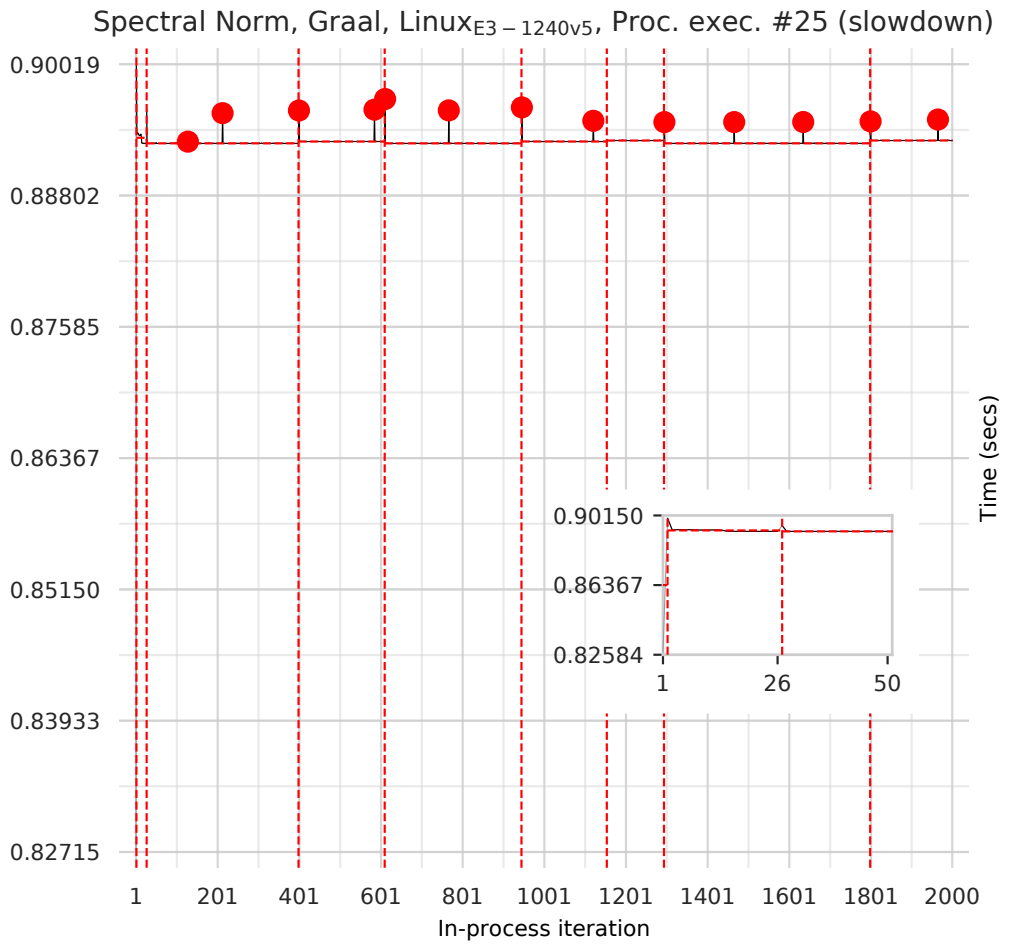

## C.4 Examples of No Steady State Behaviour

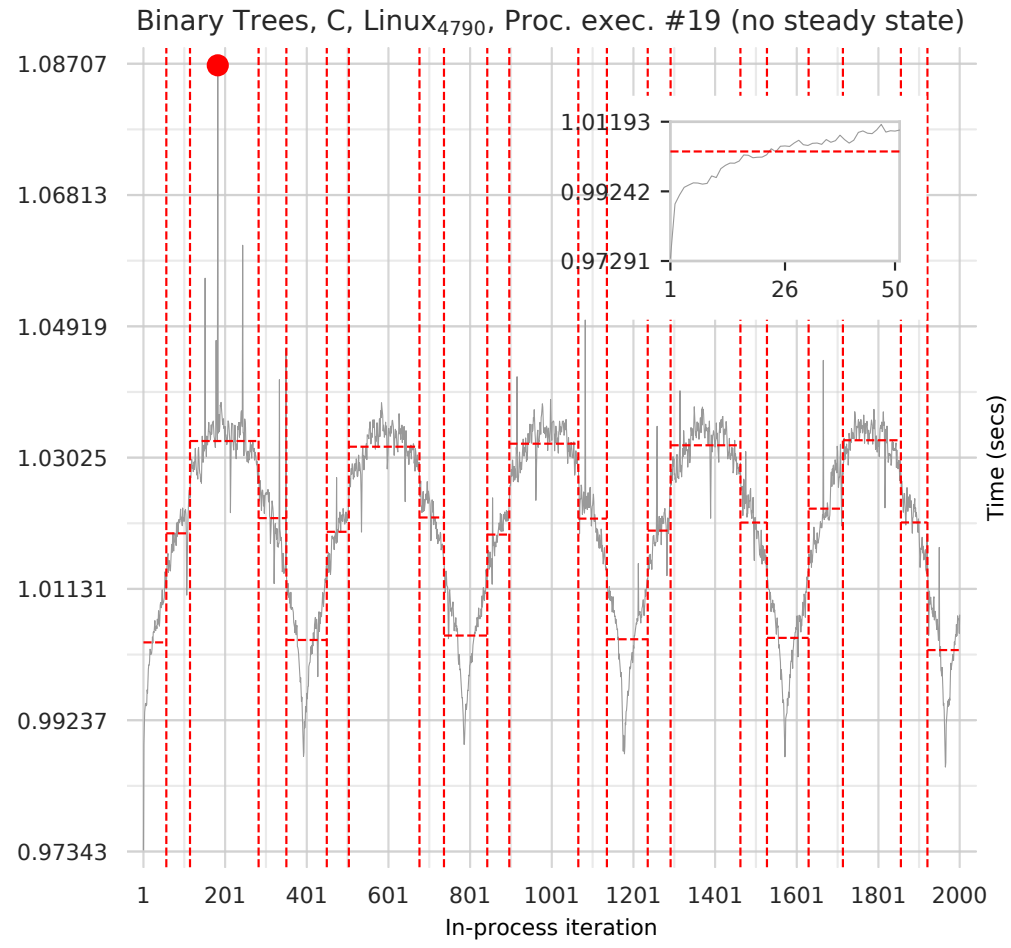

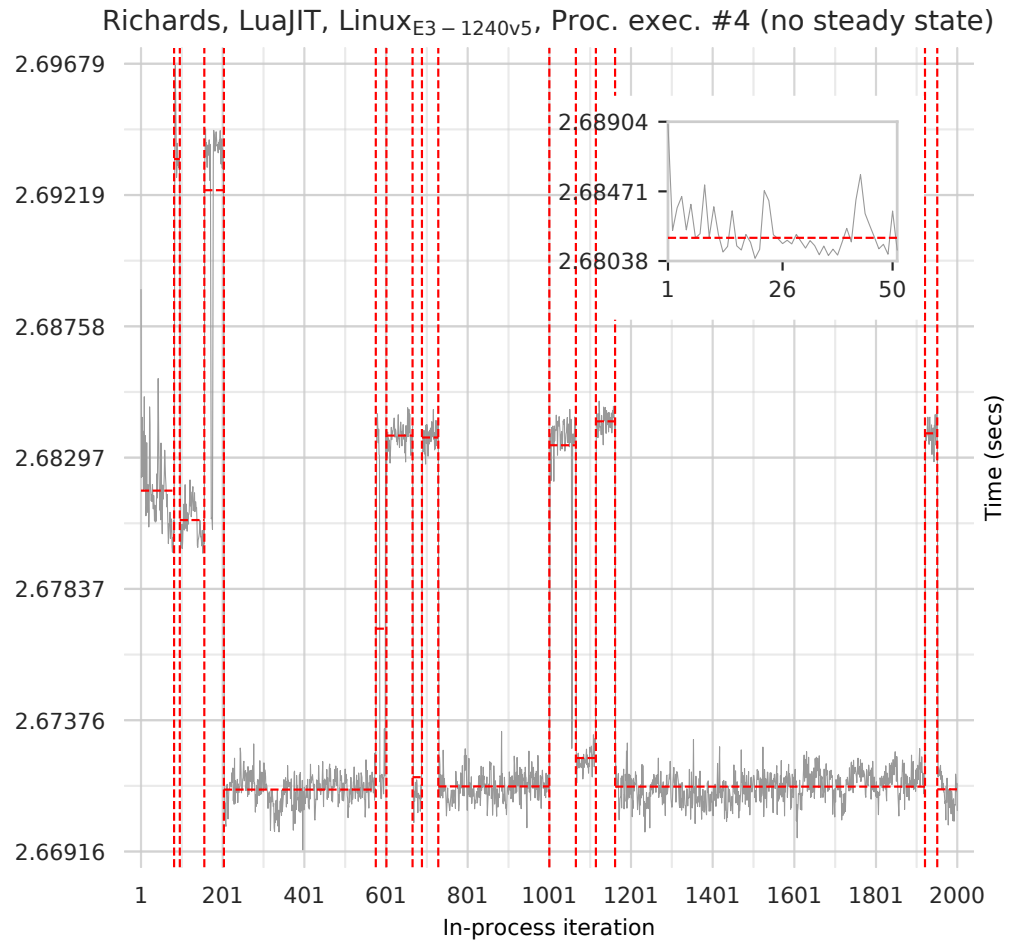

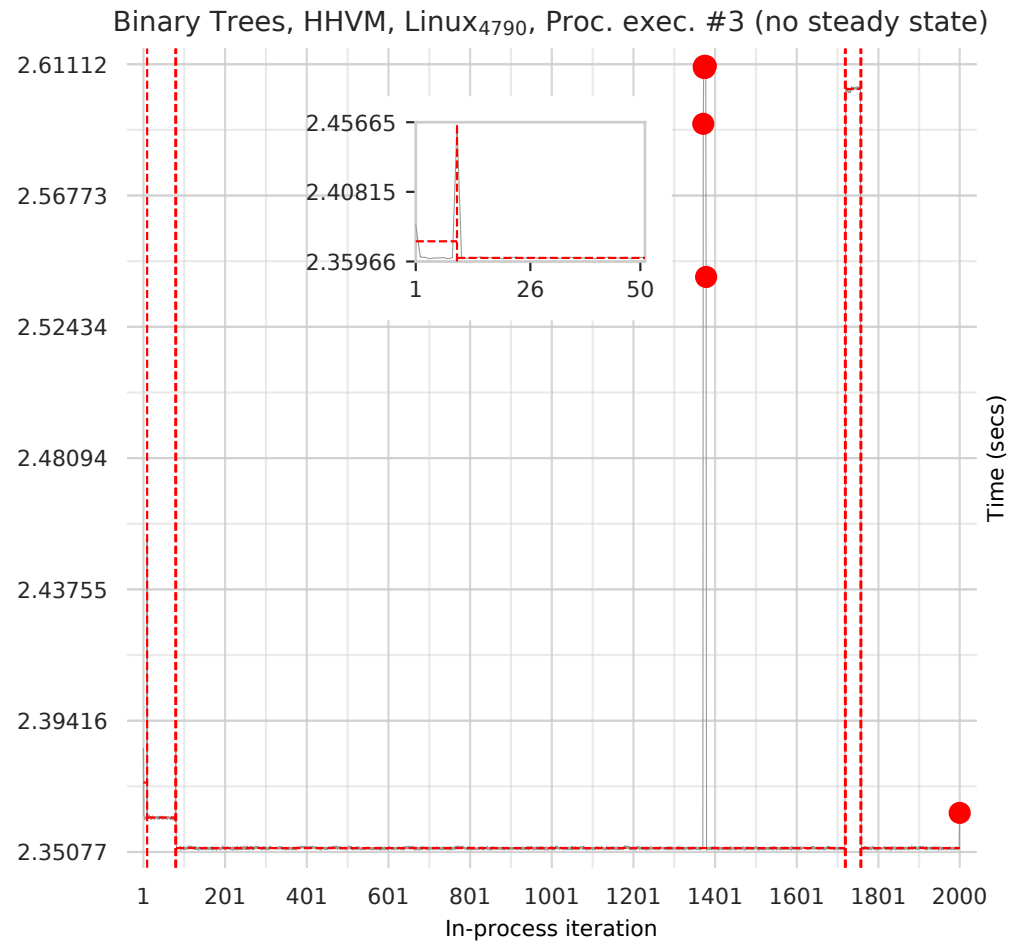

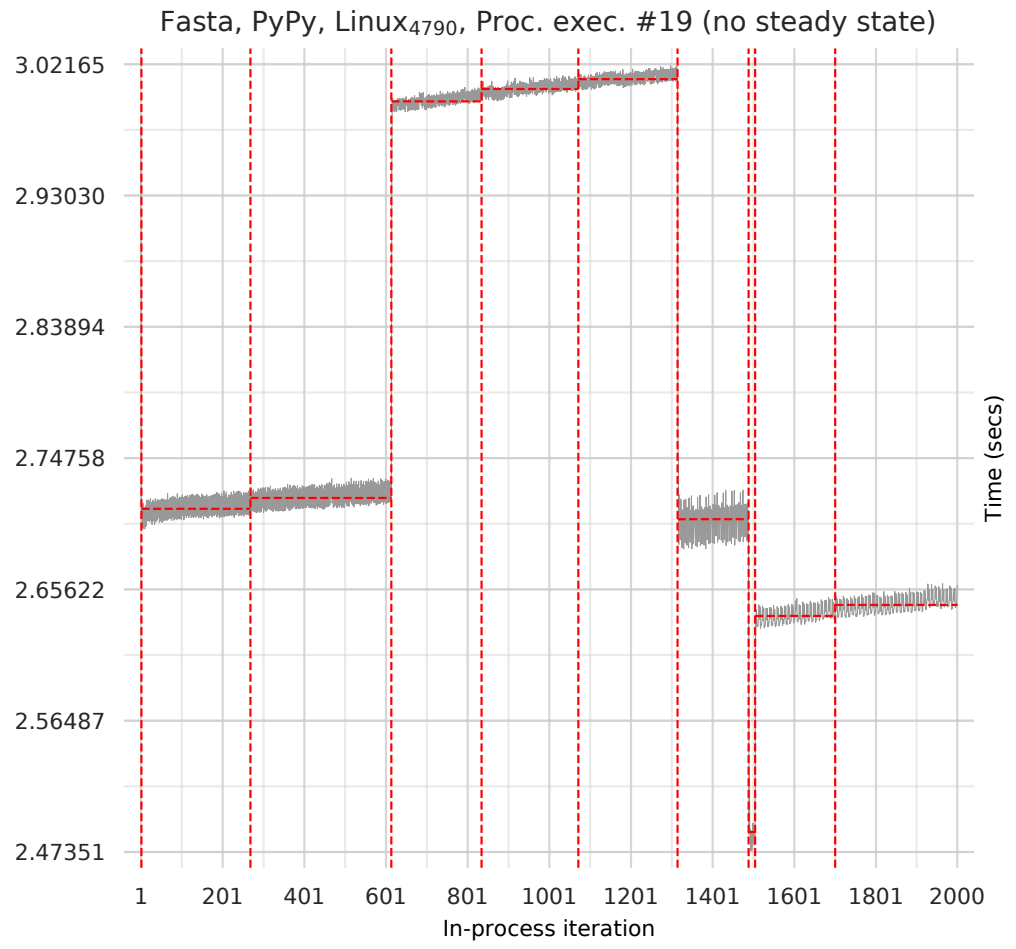